
\message{Defining fonts...Sun version 1/28/88 }
%
%
%
\font\twelverm=cmr12
\font\elevenrm=cmr10 scaled 1095
\font\tenrm=cmr10
\font\ninerm=cmr9

\font\sevenrm=cmr7

\font\fiverm=cmr5

%
\font\twelvei=cmmi12
\font\eleveni=cmmi10 scaled 1095
\font\teni=cmmi10
\font\ninei=cmmi9

\font\seveni=cmmi7

\font\fivei=cmmi5

%
\font\twelvesy=cmsy10 scaled 1200
\font\elevensy=cmsy10 scaled 1095
\font\tensy=cmsy10
\font\ninesy=cmsy9

\font\sevensy=cmsy7

\font\fivesy=cmsy5

%

%

%

%

%
\font\twelvebf=cmbx12
\font\elevenbf=cmbx10 scaled 1095
\font\tenbf=cmbx10
\font\ninebf=cmbx9

\font\sevenbf=cmbx7

\font\fivebf=cmbx5

%
\font\twelvett=cmtt10 scaled 1200
\font\eleventt=cmtt10 scaled 1095
\font\tentt=cmtt10

%

%
\font\twelvesl=cmsl12
\font\elevensl=cmsl10 scaled 1095
\font\tensl=cmsl10

%
\font\twelveit=cmti12
\font\elevenit=cmti10 scaled 1095
\font\tenit=cmti10

%

%

%

%

\font\elevenb=cmb10 scaled 1095
\font\tenb=cmb10

%

%

\def\tenpoint{
	\baselineskip11pt
	\setbox\strutbox=\hbox{\vrule height8pt depth3pt width0pt}%
	\textfont0=\tenrm \scriptfont0=\sevenrm \scriptscriptfont0=\fiverm%
	\def\rm{\fam0 \tenrm}
	\textfont1=\teni \scriptfont1=\seveni \scriptscriptfont1=\fivei%

	\textfont2=\tensy \scriptfont2=\sevensy \scriptscriptfont2=\fivesy%

	\def\it{\fam\itfam\tenit}
	\textfont\itfam=\tenit
	\def\sl{\fam\slfam\tensl} 
	\textfont\slfam=\tensl
	\def\bf{\fam\bffam\tenbf} 
	\textfont\bffam=\tenbf \scriptfont\bffam=\sevenbf
	\scriptscriptfont\bffam=\fivebf
	\def\tt{\fam\ttfam\tentt} 
	\textfont\ttfam=\tentt

	\def\emp##1{{\tenb ##1}}%
	\rm
}

\def\elevenpoint{
	\baselineskip13pt
	\setbox\strutbox=\hbox{\vrule height8pt depth3pt width0pt}%
	\textfont0=\elevenrm \scriptfont0=\sevenrm \scriptscriptfont0=\fiverm%
	\def\rm{\fam0 \elevenrm}
	\textfont1=\eleveni \scriptfont1=\seveni \scriptscriptfont1=\fivei%

	\textfont2=\elevensy \scriptfont2=\sevensy \scriptscriptfont2=\fivesy%

	\def\it{\fam\itfam\elevenit}
	\textfont\itfam=\elevenit
	\def\sl{\fam\slfam\elevensl} 
	\textfont\slfam=\elevensl
	\def\bf{\fam\bffam\elevenbf} 
	\textfont\bffam=\elevenbf \scriptfont\bffam=\sevenbf
	\scriptscriptfont\bffam=\fivebf
	\def\tt{\fam\ttfam\eleventt} 
	\textfont\ttfam=\eleventt

	\def\emp##1{{\elevenb ##1}}%
	\rm
}

\def\twelvepoint{
	\baselineskip14pt
	\setbox\strutbox=\hbox{\vrule height10pt depth4pt width0pt}%
	\textfont0=\twelverm \scriptfont0=\ninerm \scriptscriptfont0=\sevenrm%
	\def\rm{\fam0 \twelverm}
	\textfont1=\twelvei \scriptfont1=\ninei \scriptscriptfont1=\seveni%

	\textfont2=\twelvesy \scriptfont2=\ninesy \scriptscriptfont2=\sevensy%

	\def\it{\fam\itfam\twelveit}
	\textfont\itfam=\twelveit
	\def\sl{\fam\slfam\twelvesl} 
	\textfont\slfam=\twelvesl
	\def\bf{\fam\bffam\twelvebf} 
	\textfont\bffam=\twelvebf \scriptfont\bffam=\ninebf
	\scriptscriptfont\bffam=\sevenbf
	\def\tt{\fam\ttfam\twelvett} 
	\textfont\ttfam=\twelvett

	\rm
}

%
%
%
%
%


\def\subhead#1{\penalty1000\hbox to \hsize{#1\hss}\penalty1000}	





\def\spose#1{\hbox to 0pt{#1\hss}}

\newcount\notenumber
\notenumber=1
\newcount\eqnumber
\eqnumber=1
\newcount\tempnumber
\tempnumber=1
\newcount\fignumber
\fignumber=1
\newbox\abstr


\def\kpc{{\rm\,kpc}}

\def\msun{{\rm\,M_\odot}}

\def\yr{{\rm\,yr}}

\def\note#1{\footnote{$^{\the\notenumber}$}{#1}\global\advance\notenumber by 1}
\def\foot#1{\raise3pt\hbox{\tenrm \the\notenumber}
     \hfil\par\vskip3pt\hrule\vskip6pt
     \noindent\raise3pt\hbox{\tenrm \the\notenumber}
     #1\par\vskip6pt\hrule\vskip3pt\noindent\global\advance\notenumber by 1}


\def\Dt{\spose{\raise 1.5ex\hbox{\hskip3pt$\mathchar"201$}}}	
\def\dt{\spose{\raise 1.0ex\hbox{\hskip2pt$\mathchar"201$}}}	

\def\neweq{\rm\chaphead\the\eqnumber \global\advance\eqnumber by 1}
\def\refeq#1{\advance\eqnumber by-#1 \rm\chaphead\the\eqnumber
     \advance\eqnumber by #1}
\def\lasteq{\advance\eqnumber by -1 {\rm\chaphead\the\eqnumber}\advance
     \eqnumber by 1}
\def\nameq#1{\xdef#1{\chaphead\the\eqnumber}}
\def\reffeq#1#2{\tempnumber=#1 \advance\tempnumber by #2
     {\chaphead\the\tempnumber}}

\def\newfig{{\chaphead\the\fignumber}\global\advance\fignumber by 1}
\def\newfiga#1{{\chaphead\the\fignumber{#1}}\global\advance\fignumber by 1}
\def\lastfig#1{\advance\fignumber by -1 {\chaphead\the\fignumber{#1}}
     \advance\fignumber by 1}
\def\reffig#1{\advance\fignumber by -#1 {\chaphead\the\fignumber }
     \advance\fignumber by #1}
\def\namfig#1{\xdef#1{\chaphead\the\fignumber}}
\def\refindent{\par\noindent\parskip=4pt\hangindent=3pc\hangafter=1 }
\def\apj#1/#2/#3/{\refindent#1.  {\sl Ap.\/J.}, {\bf#2}, #3.}
\def\apjlett#1/#2/#3/{\refindent#1.  {\sl Ap.\/J. (Letters)}, {\bf#2}, #3.}
\def\apjsupp#1/#2/#3/{\refindent#1.  {\sl Ap.\/J. Suppl.}, {\bf#2}, #3.}
\def\mn#1/#2/#3/{\refindent#1.  {\sl M.\/N.\/R.\/A.\/S.\/}, {\bf#2}, #3.}
\def\mnras#1/#2/#3/{\refindent#1.  {\sl M.\/N.\/R.\/A.\/S.\/}, {\bf#2}, #3.}
\def\aj#1/#2/#3/{\refindent#1.  {\sl A.\/J. }{\bf#2}, #3.}
\def\aa#1/#2/#3/{\refindent#1.  {\sl Astr. Ap.}, {\bf#2}, #3.}
\def\aasup#1/#2/#3/{\refindent#1.  {\sl Astr. Ap. Suppl. Ser.}, {\bf#2}, #3.}
\def\Nature#1/#2/#3/{\refindent#1.  {\sl Nature }{\bf#2}, #3.}
\def\Icarus#1/#2/#3/{\refindent#1.  {\sl Icarus }{\bf#2}, #3.}
\def\refpaper#1/#2/#3/#4/{\refindent#1.  {\sl #2}, {\bf#3}, #4.}
\def\refpapera#1/#2/#3/#4/{\refindent#1  {\sl #2}, {\bf#3}, #4.}
\def\refbook#1/{\refindent#1}
\def\science#1/#2/#3/{\refindent#1. {\sl Science }{\bf#2}, #3.}
\def\pasp#1/#2/#3/{\refindent#1.  {\sl Pub. A.S.P.}, {\bf#2}, #3.}
\def\pasj#1/#2/#3/{\refindent#1.  {\sl Pub. A.S.J.}, {\bf#2}, #3.}


\def\lta{\mathrel{\spose{\lower 3pt\hbox{$\mathchar"218$}}
     \raise 2.0pt\hbox{$\mathchar"13C$}}}
\def\gta{\mathrel{\spose{\lower 3pt\hbox{$\mathchar"218$}}
     \raise 2.0pt\hbox{$\mathchar"13E$}}}


\def\chaphead{}

\def\ctr#1{\hfil#1\hfil}  

\def\linebreak{\hfil\break}
\def\page{\par\vfill\eject}
\catcode`\@=11 
\def\vfootnote#1{\insert\footins\bgroup\tenrm
  \interlinepenalty\interfootnotelinepenalty
  \splittopskip\ht\strutbox 
  \splitmaxdepth\dp\strutbox \floatingpenalty=20000
  \leftskip=0pt \rightskip=0pt \spaceskip=0pt \xspaceskip=0pt
  \textindent{#1}\footstrut\futurelet\next\fo@t}
\catcode`\@=12


\font\titlefnt=cmbx10 scaled 1440
\font\sc=cmcsc10 scaled 1200
\font\secfnt=cmbx12
\font\subfnt=cmsl12
\font\subsubfnt=cmsl10
\def\section#1\par{\vskip0pt plus.2\vsize\penalty-100 
  \vskip0pt plus-.2\vsize\bigskip\vskip\parskip
  \message{#1}\centerline{\secfnt#1}\nobreak\medskip}
\def\subsection#1\par{\vskip0pt plus.2\vsize\penalty-50 
  \vskip0pt plus-.2\vsize\bigskip\vskip\parskip
  \message{#1}\centerline{\subfnt#1}\nobreak\medskip}
\def\subsubsection#1\par{\vskip0pt plus.2\vsize\penalty-50 
  \vskip0pt plus-.2\vsize\bigskip\vskip\parskip
  \message{#1}\centerline{\subsubfnt#1}\nobreak\medskip}
\def\title#1{\vskip 24pt plus 12pt minus 12pt\tabskip0pt plus 1000pt
   \titlefnt\halign to \hsize{\ctr{##}\cr#1\crcr}\bigskip}
\def\author#1{\bigskip\tabskip0pt plus 1000pt \sc\halign to \hsize{\ctr{##}
   \cr#1 \crcr}}

\def\date#1{\bigskip\centerline{\tenrm #1}\bigskip}
\def\abstract{\section Abstract\par\leftskip=20pt\rightskip=20pt}



\def\ref{\par\noindent\hangindent 20pt\hangafter 1}
\def\bysame{\hbox to 50pt{\leaders\hrule height 2.4pt depth -2pt\hfill .\ }}

\hoffset 0in
\voffset 0in
\hsize 6.5truein
\jot=10pt

\overfullrule=0pt
\null
\twelvepoint
\baselineskip=28truept

\def\makeheadline{\vbox to0pt{\vskip-36.5pt
  \line{\vbox to8.5pt{}\the\headline}\vss}\nointerlineskip}
\headline={\ifnum\pageno<2 \hfil
	\else \hfil Polarized Scattering \hfil\fi}
\footline={\hss\twelverm\folio\hss}
\vskip 10pt
\baselineskip=22truept
\centerline{\titlefnt Polarized Scattering in the Vicinity of Galaxies}
\vskip 10pt

\centerline{Brian W. Murphy$^{1}$ and David F. Chernoff$^{1,2}$
\footnote{}{$^1$ Department of Astromony, Cornell University}
\footnote{}{$^2$ Presidential Young Investigator}}\par

\centerline{\it Center for Radiophysics and Space Research, Cornell University,
Ithaca, NY 14853}\par

\section Abstract\par

Some bright cD galaxies in cluster cooling flows have Thomson optical
depths exceeding $10^{-2}$. A few percent of their luminosity is
scattered and appears as diffuse polarized emission. We calculate the
scattering process for different geometric combinations of luminosity
sources and scattering media. We apply our results to galaxies, with
and without active nuclei, immersed in cooling flows. We model
observations of NGC~1275 and M87 (without active nuclei) in the
presence of sky and galactic background fluxes which hinder the
measurement of the scattered light at optical wavelengths.  Current
instruments are unable to detect the scattered light from such
objects.  However, when a galaxy has an active nucleus of roughly the
same luminosity as the remainder of the galaxy in V, {\it both} the
total and polarized scattered intensity should observable on large
scales (5--30~kpc), meaning intensity levels greater than 1\% of the
background level. For typical AGN and galaxy spectral distributions,
the scattering is most easily detected at short (U) wavelengths. We
point out that a number of such cases will occur. We show that the
radiation pattern from the central nuclear region can be mapped using
the scattering. We also show that the scattered light can be used to
measure inhomogeneities in the cooling flow.
\par

\vskip 10pt

{\it Subject headings}: polarization -- galaxies:clustering -- cooling
flows -- X-rays:galaxies

\page
\section I. Introduction\par

\def\etal{{\it et al. }}
A variety of observations suggest that large quantities of plasma and
dark matter form halos around galaxies.  Continuum X-ray brightness
profiles (Jones and Forman 1984) demonstrate the existence of hot,
distributed gas while X-ray line emission (Stewart, Canizares, Fabian,
and Nulsen 1984) and centrally peaked X-ray images (Canizares,
Stewart, and Fabian 1983) show that cooling of the gas occurs near the
center of some galaxies (see Fabian, Nulsen, and Canizares 1991 for a
review of cooling flows).  Optical observations of rich clusters show
substantial line emission emanating from filaments of gas in the
cluster core and/or from the central dominant galaxy (Kent and Sargent
1979, Heckman 1981, Cowie, Hu, Jenkins, and York 1983).  Deprojection
and modeling of the X-ray data (Stewart, Fabian, Jones, and Forman
1984 and others; a review is presented by Sarazin 1986) yield density
profiles, mass inflow rates, and temperature distributions. Some of
the mass accretion rates are enormous, e.g.  1000 $\msun/\yr$ for the
cooling flow of PKS~0745-191 (Fabian \etal~1985).

The inferred electron distribution implies Thomson optical depths of
$10^{-3}$--$10^{-2}$ over characteristic length scales (impact
parameters) ranging from a galaxy radius of $\sim 20$~kpc (or inner
edge of a cooling flow) up to a cluster core radius of $\sim 100$~kpc
(or outer edge of a cooling flow).  The galaxy's scattered luminosity
is roughly $0.1$--$1$\% of its total luminosity and may be observable
in some circumstances (Syunyaev 1982; Gil'fanov, Syunyaev, and
Churazov 1987).  Some galaxies with cooling flows have active nuclei,
e.g.  for a selected sample of eight radio-loud quasars, Crawford and
Fabian (1989) find that all eight reside in clusters surrounded by hot
ionized gas.  Study of the underlying light distributions from
radio-loud quasars also indicates that many are giant elliptical
galaxies (Hutchings, Crampton, and Campbell 1984) for which cooling
flows are expected.  In general, massive galaxies with active nuclei
are prime prospects for observing scattered light as suggested by
Fabian (1989).  He pointed out that a dense cooling flow around a
quasar might yield an anisotropic optical continuum by the following
mechanism.  If the quasar's optical emission is beamed and collinear
with its radio emission, then it scatters in the hot plasma of the
cooling flow and produces continuum emission aligned with the radio
axis, as has been observed (Djorgovski \etal~1987; Chambers, Miley,
and van Breugel 1987; McNamara and O'Connell 1993).  Strong, frequency
independent polarization of the scattered radiation is the
characteristic signature of the process.  In a study of six high
redshift galaxies di Serego Alighier, Cimatti, and Fosbury (1992)
found four show a high degree of linear polarization.  Because the
alignment of this polarization was perpendicular to the radio/optical
axis they concluded that the polarization is caused by light, from an
anisotropic active nucleus, scattered into the line of sight by gas
surrounding the galaxy.

For general wavelengths, Wise and Sarazin (1990) conclude that
scattering is most likely detectable in the radio.  If the scattering
is measurable, Wise and Sarazin show that it can be used as a
cosmological distance indicator as well as a probe of the physical
conditions of the cooling flow.  Recently, Wise and Sarazin (1992)
examined the effects of inhomogeneities in cooling flows and temporal
variations in the luminosity of the central source.  The scattered
radiation can provide a record of the luminosity over the last $10^5$
years and may help identify the parent population of high-luminosity
active galactic nuclei.

The drawback with observing scattering at radio wavelengths is that
Faraday rotation effectively erases the polarization.  Distinguishing
the scattered light from other diffuse emission processes may be
problematic once the polarization, the characteristic signature of the
scattering process, is lost. At optical wavelengths Faraday rotation
should be much less of a problem but detection may be more difficult.
In this paper we investigate the intensity and polarization properties
of {\it visible} light scattered by the typical, inferred electron
column densities.  In our model a bright galaxy (with or without an
active nucleus) is surrounded by a halo of ionized matter, either a
cooling flow or hot gas confined by a cluster potential. We calculate
the intensity of the scattered emission, the polarization fraction,
and the two-dimensional pattern of polarization seen on the sky. While
some polarized light is always produced in these situations, we
discuss the observational difficulties associated with its detection.
{\it We conclude that scattered, polarized light at optical
frequencies should be detectable on large scales (5--30~kpc) in
galaxies with sufficiently luminous nuclei and sufficiently dense
cooling flows.  The rough criteria are that the apparent luminosity of
the nucleus exceed that of the parent galaxy and that the Thomson
optical depth be of order $1$\%}.

Observation of scattered light probes (i) the angular distribution of
the radiation from the nucleus and (ii) the angular and radial
distribution of material that surrounds the illuminating source. The
first issue is especially relevant to theories that attempt to unify
disparate classes of active galaxies with geometric beaming arguments
(Barthel 1989).  The second issue is important because the scattered
polarized light can be used to probe length scales in cooling flows
that have hitherto been inaccessible.  X-ray observations have been
the traditional tool for studying cooling flows, and much of our
knowledge of them comes from the spectral and angular distribution
recorded by the Einstein Observatory and more recently by ROSAT
(Sarazin, O'Connell, and McNamara 1992). Here we show that scattered
light provides a complementary source of information at angular scales
smaller than the resolution of X-ray telescopes.

The plan of the paper is as follows. In \S IIa-f we review the physics
of scattering. We lay out our assumptions and approximations and solve
several simple, model problems.  In \S IIIa we show that the finite
size of the luminosity source limits the fractional polarization at
small distances from the center of the source. In \S IIIb-f we carry
out calculations for the prototypical sources M87 and NGC 1275, cD's
with active nuclei, and an atypical source, PKS~0745-191, and discuss
limitations imposed by diffuse emission from the parent galaxy and the
sky background. Readers primarily interested in applications will want
to scan \S IIa and \S IIf and then continue with \S III.  In \S IV we
summarize our work and future possibilities for research.

\section II. Physics of Scattering\par
\subsection a. Formalism\par

\def\khat{{\hat k}}
\def\nhat{{\hat n}}
\def\epo{{\hat \epsilon_1}}
\def\ept{{\hat \epsilon_2}}

We begin by assuming that a point source emits completely unpolarized
light.  Later, we will consider sources of finite extent. The
scattering is assumed to be solely due to non-relativistic electrons.
Although heavy elements will be present if the gas is not primordial,
the characteristic temperatures of cooling flows are large enough that
grains will not survive except in the regions where star formation has
begun.  We ignore a photon's typical frequency shift $\delta \nu / \nu
\simeq (kT/m_ec^2)^{1/2} = 0.13 T_8^{1/2}$, so that the radiative
transfer is frequency independent.  We also assume that the optical
depth is everywhere small, so that single-photon scattering is a good
approximation.  Finally, we assume that the luminosity of the central
source is constant in time.

The basic geometry is illustrated in Figure 1.  Our discussion follows
the standard treatment of Thomson scattering (e.g. Rybicki and
Lightman 1979). Let the source be at the origin. Let the direction of
the light emitted be $\khat$ and let the direction of the scattered
light be $\nhat$. The plane in which $\khat$ and $\nhat$ lie is the
plane determined by the position of the source, the scattering
material, and the observer. We define the first polarization vector,
$\epo$, as the direction perpendicular to $\khat$ in that plane (or
$\khat \times (\nhat \times \khat)$) and the second, $\ept$, as the
direction perpendicular to $\nhat$ and $\khat$ (or $\khat \times
\nhat$).

The differential Thomson cross section is
$$ {d \sigma_T \over d \Omega} = r_{e}^2 (1 -
	         \left( {\hat \epsilon} \cdot \nhat \right)^2 ) \; ,
\eqno(\neweq)
$$
where $r_{e}$ is the classical radius of the electron.  The
unpolarized light emitted by the source may be decomposed into two
components of equal intensity along $\epo$ and $\ept$.  The scattered
light is polarized because $\epo \cdot \nhat$ is, in general, not
equal to $\ept \cdot \nhat = 0$. Let $\theta$ be the angle between
$\khat$ and $\nhat$.  The observed light, that is the light scattered
into the line of sight, is described by two polarization directions,
each perpendicular to $\nhat$. We refer to the components
perpendicular and parallel to $\khat$ by $\perp$ and $\parallel$,
respectively. The ``emissivities'' for each mode at position $\vec s$
are
\def\jper{{j_{\perp}}}
\def\jpar{{j_{\parallel}}}
\def\svec{{\vec s}}
$$
\jper(\svec) = \left( F \over 2 \right) r_{e}^2 n_e
$$
$$
\jpar(\svec) = \left( F \over 2 \right) r_{e}^2 n_e\cos^2 \theta \; ,
\eqno(\neweq)
$$
where $F$ is the flux from the source and $n_e$ is the electron density.

The intensity seen by the observer is the integral over the line of
sight ($l$) of the scattered emission in each polarization mode.  The
two polarization directions are defined by
$$
{\hat e_{\perp}} = {\hat \epsilon_1} \times {\hat n}
$$
$$
{\hat e_{\parallel}} = {\hat e_{\perp}} \times {\hat n} \; .
\eqno(\neweq)
$$
(In the quadrupole tensor, the signs of $\hat e_{\perp}$ and $\hat
e_{\parallel}$ are irrelevant.) Figure 2 illustrates the polarization
directions as viewed by the observer. They lie along the
``tangential'' (${\hat e_{\perp}}$) and ``radial'' (${\hat
e_{\parallel}}$) directions of a circle whose center coincides with the
luminosity source. They are the same at each point along a fixed line
of sight. The intensities are
\nameq{\Ieqns}
\def\Iper{{I_{\perp}}}
\def\Ipar{{I_{\parallel}}}
\def\intii{{\int_{-\infty}^{\infty}}}
$$
\Iper = \intii \jper(\svec) dl = {3L \sigma_T \over 64 \pi^2}
        \intii {n_e(\svec) \over s^2} dl
$$
$$
\Ipar = \intii \jpar(\svec) dl = {3L \sigma_T \over 64 \pi^2}
        \intii {n_e(\svec) l^2 \over s^4} dl \; ,
\eqno(\neweq)
$$
where $L$ is the luminosity of the source. The fraction of
polarized light is
\nameq{\Pieqn}
$$
\Pi = \left|{\Iper - \Ipar \over \Iper + \Ipar}\right| \; .
\eqno(\neweq)
$$
For a single source the observed plane of polarization lies along
${\hat e_{\perp}}$.

\subsection b. Point Source with Centered Power-Law Gas Distribution\par

Now we consider a very simple scattering situation, a power-law
electron distribution
\nameq{\powerlaweqn}
$$
n_e(r) = n_{c} \left( r_c \over r \right)^\alpha \; ,
\eqno(\neweq)
$$
centered on the luminosity source.
The total intensity of the scattered light at impact parameter $h$ is
\nameq{\Itot}
$$
I = L {n_{c} \sigma_T \over h} \left( {r_c \over h} \right)^{\alpha}
{3 (\alpha +3) \Gamma({\alpha + 1 \over 2}) \over 128 \pi^{3/2}
\Gamma({\alpha + 4 \over 2} )} \;  .
\eqno(\neweq)
$$
Here, $\Gamma$ is the Gamma function.  The intensity of the
scattered light is of course dependent on the flux of the source and
the optical depth at the distance $h$.  This expression for the total
scattered intensity is identical to eq. (2.2) of Wise and Sarazin
(1992) modulo a difference in units of $4\pi$. We see that $I\sim F(h)
\tau(h)$ where $F(h) \sim$ flux and $\tau(h)$ is the optical depth
$\sim (n_c\sigma_T h) (r_c/h)^{\alpha}$. The fraction of polarized
light is
\def\intpp{{ \int_{-\pi/2}^{\pi/2}}}
\nameq{\Piforpowerlaw}
$$
\Pi = { \intpp \cos^{2+\alpha}(\theta) d\theta \over
        \intpp (1 + \sin^2(\theta)) \cos^{2+\alpha}(\theta) d\theta  }
= {1 + \alpha \over 3 + \alpha } \; .
\eqno(\neweq)
$$
For a broad range of power laws a significant fraction of the
scattered light is polarized.

The observer measures the Stokes parameters relative to some
sky coordinate system ($\hat x$, $\hat y$) (see Figure 2) of his choice.
 From our results above, the scattering from a point source is
\nameq{\IQUV}
$$
I = \Iper + \Ipar
$$
$$
Q = (\Iper - \Ipar) \cos 2 \Psi
$$
$$
U = (\Iper - \Ipar) \sin 2 \Psi
$$
$$
V = 0 \; ,
\eqno(\neweq)
$$
where $\Psi$ is the angle between ${\hat x}$ and ${\hat e_{\perp}}$. For
multiple sources of quasi-monochromatic radiation ($\Iper_i$,
$\Ipar_i$, and $\Psi_i$), the Stokes parameters add up to give $(I_{tot},
Q_{tot}, U_{tot}, V_{tot})$. The total intensity seen at a point is
$I_{tot}$, the total polarization is
\nameq{\PITOT}
$$
\Pi_{tot} = { \sqrt{ Q_{tot}^2 + U_{tot}^2 + V_{tot}^2 } \over I_{tot} }
\; , \eqno(\neweq)
$$
and the angle of polarization is given by
\nameq{\Angle}
$$
\tan 2 \Psi = { U_{tot} \over Q_{tot} } .
\eqno(\neweq)
$$

\subsection c. Point Source with Off-Center Gas Distribution\par

\subsubsection i.) Analytic Limiting Forms\par

In general, the gas responsible for the scattering is hardly likely to
be symmetrically distributed about the illuminating source.  In this
section we relax the assumption of symmetry. We illustrate how
displacing the luminosity center from the gas center alters the
polarization fraction, the intensity of the scattered light, and the
geometry of the polarization pattern. Assume an electron distribution
with a scale--free power--law profile ({\it cf.} eq. ({\powerlaweqn}))
with $\alpha=1$) centered at the point ${\vec r_0} = (x_0, y_0, z_0)$
and a luminosity point source at the origin.  Without loss of
generality, we assume that the observer views the scattered light
along the $\hat z$ direction.  We derive limiting forms for the
scattering and polarization for lines of sight passing near $\vec r_0$
and the origin, where the scattered intensity has local maxima.
Elsewhere, the paucity of electrons and/or the dilution of flux from
the source decreases the scattered intensity.  We provide a systematic
expansion in angular harmonics that describes the asymptotic behavior
far from the center.

Let the impact parameter of the line of sight with respect to the
origin be $p = \sqrt{ x^2 + y^2 }$ and with respect to the density
peak be $q = \sqrt{ (x-x_0)^2 + (y-y_0)^2 }$. Letting $I_0= ${\hfill\break}$3 L
n_{c} \sigma_T r_c / 64 \pi^2$, eq. (\Ieqns) becomes
$$
{ \Iper \choose \Ipar } = I_0 \int_{-\infty}^{\infty} dl { 1
\choose {l^2 \over p^2 + l^2} } {1 \over (p^2 + l^2) (q^2 + (l-z_0)^2
)^{1/2} } .
\eqno(\neweq)
$$
First, for lines of sight that pass near the density peak ($q/z_0 \to
0$ and finite $p/z_0$), we expand in powers of $(q/z_0)$ and integrate
term by term to find
\nameq{\Idenmax}
$$
\Iper \to
      { I_0 \over (z_0^2 + p^2) }
           \left(
             -2 {\rm ln} \left( q \over z_0 \right)
             + {\rm ln} 4 \left(1  + \left( p \over z_0 \right)^2 \right)
             - {2 \tan^{-1}\left( p \over z_0 \right) \over
                           \left( p \over z_0 \right) }
                           + O\left( q \over z_0 \right)
           \right)
$$
$$
\Ipar  \to
       { I_0 z_0^2 \over (z_0^2 + p^2)^2 }
        \left( -2 {\rm ln} \left( q \over z_0 \right)
         + {\rm ln} 4 \left(1  + \left( p \over z_0 \right)^2 \right) \right.
$$
$$              +
                  \left. \left( \left( p \over z_0 \right)^2 - 1 \right)
                      \left({\tan^{-1} \left( p \over z_0 \right) \over
                           \left( p \over z_0 \right) }  + 1 \right)
                           + O\left( q \over z_0 \right) \right) .
\eqno(\neweq)
$$
Formally, $\Iper$ and $\Ipar$ diverge logarithmically near the density
peak, although one (or both) of the assumptions, optically thin
radiative transfer and  unbounded growth in density near $\vec r_0$,
must break down.  As $q \to 0$ the fractional polarization
approaches the limit
\nameq{\poldenmax}
$$
\Pi \to {p^2 \over 2z_0^2 + p^2} \to
        {r_0^2 - z_0^2 \over r_0^2 + z_0^2} .
\eqno(\neweq)
$$
When the peak and luminosity source lie in the same plane ($p/z_0 \to
\infty$), the fractional polarization is a maximum.  At a fixed-impact
parameter, a larger line-of-sight separation lowers $\Pi$. To
summarize, {\it significant fractional polarization is expected near a
density peak whose three-dimensional displacement from the luminosity
source is less than a modest multiple of the impact parameter.}

A systematic expansion for scattering at large impact parameters may
be developed as follows. We express the density field (centered at
$\vec r_0$ with the form $1/| {\vec r} - {\vec r_0}|$) in spherical
harmonics and powers of the distance from the origin. The integral
over the line of sight is converted to an angular integral and carried
out term by term.  Let $(r_0, \theta_0, \phi_0)$ be the spherical
coordinates of the density field center, measured with respect to the
origin (the luminosity point source). The impact parameter of the line
of sight in the $\hat z$ direction is $p$ at angle $\phi$.  For $p >
r_0$ we have the exact solution $$ {\Iper \choose \Ipar} = I_0
\sum_{\scriptstyle l=0...\infty
\atop{\scriptstyle |m| \le l
\atop\scriptstyle m+l {\rm \ \ even}}}
{A_{lm} \choose B_{lm}}
e^{i m (\phi - \phi_0)} P_l^m( \cos \theta_0 ) {r^l_0 \over p^{l+2}}
\; , \eqno(\neweq)
$$
where
$$
A_{lm} = 2 \left(-1\right)^{(m + l)/2} { (l-m)! \over (2l + 1) ! ! }
$$
$$
B_{lm} = \left( 1 - { (l-m+2)(l+m+2) \over 2 (2l+3) } \right) A_{lm} .
\eqno(\neweq)
$$
To order $l=2$ these expansions yield the following explicit forms:
\nameq{\Iradmax}
$$
{\Iper \over I_0} =
{2 \over p^2} +
{4 r_0 \over 3 p^3} \sin \theta_0 \cos( \phi - \phi_0 ) +
{4 r_0^2 \over 5 p^4}
   \left( \sin^2 \theta_0 \cos( 2(\phi - \phi_0) ) -
          {1 \over 6} \left( 3 \cos^2 \theta_0 - 1 \right) \right)
$$
$$
{\Ipar \over I_0} =
{2 \over 3 p^2} +
{4 r_0 \over 15 p^3} \sin \theta_0 \cos( \phi - \phi_0 ) +
{4 r_0^2 \over 35 p^4}
   \left( \sin^2 \theta_0 \cos( 2(\phi - \phi_0) ) +
          {1 \over 6} \left( 3 \cos^2 \theta_0 - 1 \right) \right) .
\eqno(\neweq)
$$
The fractional polarization at large distances (to order
$l=1$) is
\nameq{\polradmax}
$$
\Pi \to {1 \over 2} + {r_0 \over 10 p} \sin \theta_0 \cos( \phi - \phi_0 )
\; . \eqno(\neweq)
$$
To summarize, {\it at large distances from the luminosity source and
density peak, the scattering approaches the simple form for a point
source with a centered power law ({\it cf.} \S IIb).} As $p \to \infty$, $\Pi$
agrees with eq.  (\Piforpowerlaw) for the adopted density power law
($\alpha =1$).

The term by term integration may also be carried out for lines of
sight that pass within the spherical shell of radius $p < r_0$, but in
this case there are three separate regions of integration and the
formulae are more complicated. A consistent expansion is most easily
derived using cylindrical coordinates. To summarize the steps: expand
$1/|{\vec r - \vec r_0}|$ using modified Bessel functions ($I_m$ and
$K_m$), perform the line-of-sight integral in $z$, and then substitute
a series expansion for $I_m$ (8.445 of Gradshteyn and Ryzhik 1980;
hereafter G\&R) and an integral definition for $K_m$ (G\&R 8.432.3).
An asymptotic expansion of the integrand gives integrals which may be
reduced term by term to Beta functions. The results can be compactly
expressed using $s =\sqrt{ (x^2 + y^2)/(x_0^2 + y_0^2) }$ and $t
=\sqrt{ z_0^2 /(x_0^2 + y_0^2) }$; we assume the ordering $s$, $t$ $<
1$. This ordering means that (i) the line of sight passes much closer
to the luminosity source than the density peak and (ii) the density
peak's displacement along the line of sight is less than or equal to
its projected separation (other orderings may also be analogously
treated). Let
$$
\eqalignno{
F_m(s,t,\delta) &= \sum_{k=0}^{\infty} \left( s^{m + 2k + 1 + \delta}
         \over k ! (m+k) ! \right)
                \sum_{n=0}^{\infty} { (-2)^n \over n ! }\cr
&\qquad
       { \Gamma(m + k + {n + 1 + \delta \over 2})
         \Gamma(    k + {n + 1 + \delta \over 2})
       }
             \left(  (s - i t)^n + (s + i t)^n \right)
&(\neweq)\cr}
$$
and let $a_i = 1$ for $i >0$ and $a_0 = 1/2$:
$$
{\Iper \choose \Ipar} =
{I_0 \over p^2 } \sum_{m=0}^{\infty} a_m \cos{m(\phi-\phi_0)}
{F_m(s,t,0) \choose {F_m(s,t,0) \over 2} - F_m(s,t,1) } .
\eqno(\neweq)
$$

Using the above expressions it is straightforward to give $\Ipar$ and
$\Iper$ to order $s^{\mu_s} t^{\mu_t}$ for $\mu_s + \mu_t < M$ for a given
$M$. For $M=2$ the result is independent of $t$,
\nameq{\Ilummax}
$$
{\Iper \over I_0} =
{s \pi \over p^2} \left( 1 - {2 \over \pi} s +
         s \cos(\phi-\phi_0) \right)
\eqno(\neweq)
$$
and
$$
{\Ipar \over I_0} =
{s \pi \over 2 p^2} \left( 1 - {4 \over \pi} s +
         s \cos(\phi-\phi_0) \right) .
\eqno(\neweq)
$$
The divergence in intensity as $p \to 0$ may be traced to the
assumption of a luminous point source. Note that the limiting behavior
of the fractional polarization is
\nameq{\pollummax}
$$
\Pi \to {1 \over 3} (1 + {8 s \over 3 \pi} - 2 s \cos(\phi-\phi_0))
\; ,\eqno(\neweq)
$$
which is the same as for a point source embedded in a scattering gas
of uniform density. To summarize, {\it the fractional polarization
expected in the vicinity of a luminosity point source approaches that
of a point source in a uniform density medium.}

\subsubsection ii.) Numerical Results\par

We now proceed with a more detailed description via a numerical
calculation.  We have carried out the quadratures in eqs. (\Ieqns) for
a grid of 100 by 100 lines of sight. As in the analytic work above we
take a scale-free power law (eq. (\powerlaweqn) with $\alpha=1$) for
the density distribution.  The source is located in the plane of the
gas center (co-planar geometry, $z_0 = 0$) or displaced out of the
plane ($\sqrt{x_0^2 + y_0^2}/z_0 = 1/4$).  Figure 3a shows the
polarization fraction (length of the line segments), the angle of
polarization (direction of the line segments), and the intensity of
the scattered light (contours) for the co-planar geometry. Figure 3a
and future figures are centered about the gas-density maximum ($\vec
r_0$). The maximum intensity occurs near the source while a local
maximum occurs near the gas center.  The numerically derived
fractional polarization, shown in Figure 3b, is a minimum near the
source ($\sim$0.35; compare eq. (\pollummax)) and a maximum near the
gas center ($\sim$0.75; compare eq. (\poldenmax)). In the vicinity of
the gas center and the point source, the polarization ranges between
these extremes; at large radii it approaches 0.5, as would be expected
for a point source embedded in a gas following $r^{-1}$ power law
(eqs.  (\Piforpowerlaw) or (\polradmax)).  Of course, at the gas
center the numerically determined fractional polarization should
approach its maximum value of unity, while at the source it should go
to 1/3.  These limits are only approximately observed in our numerical
model because of the finite resolution in the 100 by 100 grid. When
the source and gas center are displaced out of the plane (Figure 3c)
scattering near the gas center is diminished (the flux from the source
at the gas center is less; compare eq.  (\Idenmax)) and the
polarization decreases (the scattering angle, $\theta$, becomes small
at the high-density region; compare eq.  (\poldenmax)).

\subsection d. Comparison of Center and Off-Center Gas Distributions\par

We now consider the scattering when both a centered and off-centered
distribution are present. We want to derive simple criterion to
distinguish which type of scattering dominates.  Of course, since the
radiative transfer is linear in our approximation the individual
results may be simply added to give $I_{tot}$ and $\Pi_{tot}$ as
described in \S IIb. First, assume both distributions lie in the same plane
and that the off-center distribution is a power law with $\alpha=1$.
The intensity from the off-center ``clump'' is given by
$$
I_{\perp}^{cl} = {I_0 \over (1-u^2)^{1/2} p^2} {\rm ln} \left[{1 + \sqrt{1-u^2}
\over 1 - \sqrt{1-u^2}} \right]
$$
$$
I_{\parallel}^{cl} = {I_0 \over (1-u^2)^{3/2} p^2} \left( \sqrt{1-u^2} -
{u^2 \over 2} {\rm ln}  \left[{1 + \sqrt{1-u^2}
\over 1 - \sqrt{1-u^2}} \right] \right)
\eqno(\neweq)
$$
where $u=q/p$, and $q$, $p$, and $I_0$ have the same definitions as
above.  In the limit that the clump is centered on the galaxy, $u \to
1$ results in eqs. (\Itot) and (\Piforpowerlaw) are retrieved.

For comparison, the ratio of $I^{cl} = I^{cl}_{\perp} +
I^{cl}_{\parallel}$, the total scattered intensity from the off-center
clump, to $I^{gal}$, the intensity from an assumed galactic halo given
by eq. (\Itot), at the same impact parameter, is
$$
{I^{cl} \over I^{gal}} = {3 f(u) \over 8}
\left({n_c^{cl} r_c^{cl}
\over n_c^{gal} r_c^{gal} }\right)
\eqno(\neweq)
$$
where $f(u)$ is a function which diverges logarithmically at small
$u$ (this ignores the core radius of the clump). $n_c^{gal}$,
$n_c^{cl}$, $r_c^{gal}$, and $r_c^{cl}$ are the scaling densities and
radii used in eq. (6) for the off-center clump and the galaxy's gas
distribution at the center.  For fixed column densities of the clump
and galaxy, the ratio depends only upon $f(u)$.  At a fixed distance
 from the clump center (e.g. its core radius), $u \to 0$ as the clump
displacement <from the galaxy center increases.  The result suggests
that the clump scattering dominates halo scattering at large galactic
radii but is sensitive to the large extent assumed for the clump
($\alpha=1$).

It is straightforward to evaluate $I^{cl}/I^{gal}$ for arbitrary power
laws for galaxy ($\alpha^{gal} \ge 0$) and clump ($\alpha^{cl} \ge 2$)
when the line-of-sight contribution to scattering is dominated by the
points of closest approach. We find the intensity ratio scales as
$$
{I^{cl} \over I^{gal}} \approx
\left({n_c^{cl} r_c^{cl}
\over n_c^{gal} r_c^{gal} }\right)
\left( p \over r_c^{gal} \right)^{\alpha^{gal} - 1}
\left( r_c^{cl} \over  q \right)^{\alpha^{cl} - 1} .
\eqno(\neweq)
$$
For the massive cD galaxies with which we are concerned,
$\alpha^{gal} \approx 1$. Thus, for a sufficiently concentrated clump
profile ($\alpha^{cl} \ge 2$), we see that at fixed distance from the
clump center, the ratio of column densities fixes the ratio of the
brightnesses. To summarize, {\it our comparison shows that the most
significant factor dictating the clump-to-halo brightness ratio is
just the column density ratio.} If the off-center clump has an
extended ($\alpha=1$) distribution, the brightness ratio increases
slowly (logarithmically) as it is displaced from the luminosity
center.  For concentrated clumps, the ratio is insensitive to the
displacement.  We note that we have not taken account into the
extended nature of the galaxy.  Had we compared the off-center clump's
scattered intensity to the scattered intensity of an extended galaxy
instead of a point source then the ratio would have been larger, {\it
c.f.} \S IIIa.

\subsection e. Multiple Point Luminosity Sources\par

In this section we examine the polarization pattern produced by the
combination of two point sources embedded in a scale--free power--law
scattering distribution ($\alpha = 1$ in eq. (\powerlaweqn)).  For
illustrative purposes, we take the two sources to be equal in
luminosity and equidistant from the gas center as seen in projection.
One source remains fixed in the plane of the sky that passes through
the gas center, and the other source is placed at predetermined
intervals perpendicular to this plane.  In this example, the sources
are oriented so that one lies at an angle of $\pi/2$ and the other at
an angle of $0$ with respect to the gas center.

The total polarization at any given point can be calculated
numerically using eqs. (\Ieqns), (\IQUV), and (\PITOT). First, the
integrals in equations (\Ieqns) are numerically evaluated for a grid
of $100\times100$ lines of sight for a single source.  Then, the
contribution of a second source equivalent to the first but rotated by
$\pi/2$ is calculated and combined pointwise on the grid.  Figure 4a
shows the fractional polarization, the angle of polarization, and the
intensity of scattered light for a configuration with both point
sources lying in the plane of the gas center.  As expected, the
intensity has local maxima at the sources and at the center of the
scattering distribution. The new feature, however, is that there are
two regions of destructive interference seen as lobes of low
polarization. It is easy to understand their cause.  Consider the
right triangle defined by the line connecting the two sources and the
angles $\pi/4$ at each source. Clearly, near the triangle's third
vertex the polarization angle of one source is perpendicular to the
other and the flux from the sources is identical, so the scattered
polarized intensity is zero. On the other hand, the largest
polarization must occur midway on the straight line connecting the two
sources and, also, at large distances from both. In both of these
regions the polarization adds constructively and approaches the
asymptotic value of 50\% for $\alpha = 1$. We conclude that multiple
sources complicate the polarization pattern where the scattering
intensities are roughly equivalent.

The remaining plot, Figure 4b, shows the polarization pattern when the
source at $\pi/2$ (the ``secondary source'') is moved along the line
of sight to a distance four times its projected distance from the gas
center. As the source is moved out of the plane of the sky the
scattering properties undergo noticeable changes.  (It does not
matter, however, whether the source is moved into the foreground or
the background; both directions will show the same change.)  First,
the intensity of the scattered light around the secondary source
decreases from its maximum value in the plane (which is consistent
with the previous example of a single source).  Second, the fractional
polarization in the vicinity of the secondary source drops in this
example from roughly 35\% to 20\% as line of sight distance goes to
its maximum value (which is also consistent with our previous example
of a single source). Finally, the area of interference of the two
sources (the two low-polarization lobes) approaches the secondary
source as it moves out of the plane.  The lobes occur where both
galaxies provide the same flux so that the parallel and perpendicular
components of the light scattered by both sources negate one another.
That projected location must be closer to the secondary source where
it is moved out of the plane of the sky.

\subsection f. Summary\par

Our examples show that the geometry of the source and of the electron
distribution governs the polarization patterns in a simply understood
fashion. One can envision many potential applications.  It may be
possible to constrain the three-dimensional position of luminosity
sources by studying such patterns.  For example, consider a luminous
galaxy located off-center in a spherical-cluster potential that is
filled with ionized gas.  The galaxy's offset position along the line
of sight with respect to the cluster center can be determined if
enough of the light scattered by the cluster gas can be detected and
mapped. Another potential application is to use scattering
measurements to locate, identify and map inhomogeneities, {\it e.g.}
as gas becomes thermally unstable in a cooling flow. Finally, in a
cluster with many galaxies present, the constructive and destructive
effects in the polarized scattering lead to complicated
neighbor-neighbor interactions.  Detailed polarization maps may be
used to model the three-dimensional distribution of galaxies in a
cluster.  A prime candidate is the cluster 2A~0335+096 (Sarazin,
O'Connell, and McNamara 1992).  This cluster has a central D galaxy
with a bright companion nucleus at a radius of 6~kpc.  Added to this
are two other galaxies located at distances of 40~kpc.  Throughout
this region, ROSAT observations indicate that the gas density is
slightly less than $0.1 {\rm cm}^{-3}$. In the section that follows we
consider additional physical effects that complicate the use of
polarization as a potential observational tool.

\section III. Estimates of Scattering From Actual Sources

\subsection a. Extended Luminosity Sources \par

We have previously considered the intensity of the polarized
scattering, the angle of polarization, and the fractional polarization
produced by point-like luminosity sources.  In practice the luminosity
source's spatial extent may also be important, e.g.  if galactic
emission is scattered by gas within a few core radii of the galaxy
itself. In this section we consider the continuum limit for the
luminosity source and characterize the differences in the scattered
emission for a point source and an extended source.

Let $\rho_L$ be the luminosity density and $n_e$ be the scattering
density, each assumed to be spherically symmetric about the origin.
Without loss of generality, choose the line of sight in the $\hat z$
direction and $p {\hat y}$ as the impact parameter.  Let $l$ be the
distance along the line of sight and $s$ the distance between
$(x,y,z)$ and $(0,p,l)$. The scattered intensity and polarized
intensity are then given by integrals over the volume and along the
line of sight:
\nameq{\extone}
$$
\left[\matrix{I \cr P}\right]  = {3 \sigma_T \over 64 \pi^2} \int dxdydzdl
{\rho_L(x,y,z)n_e(l) \over s^4} \left[\matrix{s^2 + (l-z)^2 \cr
(y-p)^2 - x^2}\right] ,
\eqno(\neweq)
$$
where $P = \sqrt{Q^2 + U^2}$.

We assume a spatial luminosity density of the form
\nameq{\Lden}
$$
\rho_L(r) = \rho_L(0) (1 + r^2/r_c^2)^\beta  \; ,
\eqno(\neweq)
$$
where $\rho_L(0)$ is the central luminosity density, and $r_{c}$ is
the core radius.  As a specific example, we choose $\beta = -3/2$,
which implies that the surface brightness scales like $r^{-2}$ power
law for large radii.  Such a power law is commonly observed for giant
elliptical and cD galaxies (Oemler and Tinsley 1979).  The total
luminosity is dictated by the outer cutoff, which we take to be $100$
$r_c$. However, all the scattering results we discuss are essentially
independent of the cutoff because they depend upon integrals of
functions which fall much more steeply with $r$ than $\rho_L(r)$. We
assume a radial distribution of the scattering medium given by
\nameq{\nden}
$$
n_e(r) = n_{c} (1 + r^2/r_c^2)^\gamma .
\eqno(\neweq)
$$
We are primarily interested in the scattering in the halo; we
include the core $r_c$ to prevent an unrealistic divergence at the
origin (as occurs in eq. 6) As a specific example, we choose $\gamma =
-1/2$, which is roughly appropriate when the scattering medium is part
of a cooling flow (White and Sarazin 1987, Meiksin 1990), based on
model fits to X-ray observations. There are complications in using
these results to infer the scattering profile.  In the models the
X-ray intensity constrains $\langle n_e^2 \rangle$ but $\langle n_e
\rangle$ can be varied by adjusting the gas clumping.  Since the
scattered intensity is proportional to $\langle n_e \rangle$ this
yields a range in $I$, which Wise and Sarazin (1992) have shown for
the galaxy M87 can be as much as 40\%.  In addition, the loss of
ionized gas by recombination and star formation alters the inferred
scattering density.

Substituting eqs. (\Lden) and (\nden) into eq. (\extone) we find
\nameq{\exttwo}
$$
\left[\matrix{I \cr P}\right]  = {3 \sigma_T \rho_L(0) n_c r_c^2
\over 64 \pi^2} \int { d{\tilde x}d{\tilde y}d{\tilde z}d{\tilde l}
\over {\tilde s}^4 (1 + {\tilde x}^2 + {\tilde y}^2 + {\tilde z}^2)^{\beta}
(1 + {\tilde l}^2 + {\tilde p}^2)^{\gamma}}
\left[\matrix{{\tilde s}^2 + ({\tilde l} - {\tilde z})^2 \cr
({\tilde y} - {\tilde p})^2 - {\tilde x}^2}\right] \; ,
\eqno(\neweq)
$$
where the dimensionless quantities $\tilde x$, $\tilde y$, $\tilde z$,
$\tilde l$, $\tilde s$, and $\tilde p$ are $x/r_c$, $y/r_c$, etc.,
respectively. The integral has been evaluated using Romberg
integration at 20 impact parameters.

The calculated polarization as a function of $r$, $\Pi$, is shown in
Figure 5.  Because of the four-dimensional nature of the integral the
numerical computation is quite time consuming. Near the core radius up
to $10^{11}$ function evaluations per impact parameter were done.  The
relative accuracy was roughly a few percent in the inner regions. Note
that $\Pi \to 0$ as $r \to 0$, since the source illuminates the line
of sight symmetrically, and that $\Pi \to 1/2$ as $r \to \infty$
(beyond the cutoff) when the point-source approximation is good.  The
figure shows that it is necessary to be many core radii ($r > 10 \
r_c$) from the center before $\Pi$ has risen to even half its
asymptotic value.  Figure 6 shows the ratio of $I$ and $P$ for the
extended source to $I$ and $P$ for a point source.  It can be seen
that $I$ and especially $P$ are reduced significantly as the impact
parameter approaches the core of the galaxy.  Consequently, the
intensity of polarized emission is much less than would be predicted
given $I$ and the asymptotic value of $\Pi$.

Since the quadratures are time consuming, we give two empirical rules
for the total and polarized intensity as a function of radius for
extended sources,
$$
I(r) \simeq I_{pt}(r) {L_{cyl}(r) + L_{cyl}(r_{c}) \over L}
\eqno(\neweq)
$$
and
$$
P(r) \simeq {1 \over 2} I_{pt}(r) {L_{sph}(r) \over L} \; .
\eqno(\neweq)
$$
Here $L_{cyl}(r)$ and $L_{sph}(r)$ represent integrations of equation
(\Lden) giving the total luminosity within a cylinder and sphere of
projected radius $r$, respectively. $I_{pt}(r)$ is the point source
intensity ({\it c.f.} eq. (\Itot)) given by
$$
I_{pt}(r) = { L\sigma_T r_c n_{c}\over 8 \pi^2 r^2} \; .
\eqno(\neweq)
$$
Our rules of thumb given above are accurate to within 30\% for $1 \leq
r/r_c \leq 100$.  Inside the core of the galaxy the values of $I$ and
$P$ change too rapidly to be represented accurately by empirical laws.
The total luminosity of the galaxy, idealized as a point source,
provides a good fit only at large radii. For an extended source both
the scattered intensity and polarized intensity are reduced by a
significant fraction within about 10 core radii.

\subsection b.  cD Galaxies \par

In this section we predict the intensity of scattered light from cD
galaxies.  Such galaxies have two desirable properties: (i) they are
embedded in massive cooling flows; and (ii) they are usually the most
luminous galaxy in the cluster in which they reside.  A luminous
source in a dense scattering gas gives us the most favorable
conditions for observing scattered light. We examine two galaxies in
detail: M87 and NGC~1275.  Each is associated with a cooling flow and
each is the dominant galaxy in its cluster.  We assume that the
scattering gas is centered on the galaxy and that the galaxy acts as
an extended source of luminosity.  We make use of our results from
\S IIIa.

NGC~1275 has an enormous cooling flow, $\sim 300 \msun {\rm yr}^{-1}$,
and the total mass of gas lying within 200~kpc is estimated to be
$10^{13} \msun$ (Fabian {\it et al.}~1981).  We assume the luminosity
density and electron scattering gas are given by eqs. (\Lden) and (\nden),
respectively.  We take $\beta =-3/2$ and $\gamma=-1/2$ as discussed in
\S IIIa.  Integrating eq. (\nden) and expressing $n_c$ in terms of the
total mass, $M_{tot}$ within the radius $r_g$, we find
\nameq{\num}
$$
n_c = {M_{tot}
\over 2 \pi m_p r_c^3} \left( {7 \over 8} \right)
\left[{r_g\over r_c} \sqrt{{r_g^2\over r_c^2} +1} - {\rm ln}
\left({r_g\over r_c} +  \sqrt{{r_g^2\over r_c^2} +1} \right) \right]^{-1}
\; , \eqno(\neweq)
$$
where $m_p$ is the proton mass.  The 7/8 that appears in eq. (\num)
assumes the gas is composed of 75\% ionized hydrogen and 25\% doubly
ionized helium by mass.  The core radii for both scattering and
luminosity densities are assumed to be 3.2~kpc, based on the V-band
observations of Oemler (1976).  We find $n_c \sim 0.45 \; {\rm
cm}^{-3}$.  $L_V$ is known from the absolute visual magnitude of -24.27
(Oemler 1976).  Finally, we assume an outer cutoff in the luminosity
density at $r_{cut}=100r_c$ and solve for $\rho_{L_V}(0)$.  By integrating eq.
(\Lden) we find \nameq{\lumden}
$$
\rho_{L_V}(0) = {L_V \over 4\pi r^3_c}
\left[{\rm ln} \left( {r_{cut} \over r_c} + \sqrt{{r^2_{cut} \over r^2_c}
+1 }\right) - {{r_{cut} \over r_c} \over \sqrt{{r^2_{cut} \over r^2_c}
+1 }} \right]^{-1}
\eqno(\neweq)
$$
and determine $\rho_{L_V}(0) = 2.96\times 10^{-23} {\rm erg \, s^{-1}
cm^{-3}}$.  Knowing $\rho_{L_V}(0)$ and $n_c$, it is now possible to
calculate the total and polarized intensities for our model galaxy
 from eq. (\exttwo).

Figure 7a shows the total and polarized intensities, along with 1\% of
the background intensity in V. The background is composed of galaxy
plus sky emission and is a function of $r$. We include it as an
effective lower bound that the total and polarized intensities must
exceed for any hope of detection. As a practical matter, current
instruments can measure an intensity $\geq$ 1\% of the background
(Miller and Goodrich 1988). We take the sky intensity in V to be
$21.9\; {\rm mag \; arcsec}^{-2}$ (Pilachowski {\it et al.} 1989).
The galaxy's contribution to the background decreases at shorter
optical wavelengths, but the intensity of the scattered light is
similarly diminished.  As we will show in \S IIIc, when the scattered
light derives from a bluer source (such as an AGN) it is advantageous
to observe in U or B.  In this example, however, little can be gained
by doing so. An additional difficulty in making measurements of
polarization fractions below 1\% of the galaxy emission is the
grain-induced polarization that affects light traveling through the
Galaxy's interstellar medium.  Studying a large sample of nearby
stars, Serkowski, Mathewson, and Ford (1975) found an empirical
expression for the maximum polarization, $\Pi \le 0.09$ $E(B-V)$,
where $E(B-V)$ is the differential extinction.  For $|b| \le 20^{\rm
o}$ the range of $E(B-V)$ is roughly 0.01 to 0.1 (Burstein and Heiles
1982), with the lowest values found toward the poles. Hence, the
grain-induced polarization should be less than 1\% of the galaxy's
emission.  Our results show that detecting the scattered light in
NGC~1275 is currently impossible at optical wavelengths.  At no point
is $I$ above the 1\% background level.  Furthermore, the polarized
intensity is a very small fraction of the total scattered intensity.
The maximum of polarized emission occurs just beyond the core radius,
only $27.8 \; {\rm mag \; arcsec}^{-2}$, or roughly 3 magnitudes
fainter than the 1\% level of the background.

We include a second example with a different core radius and
scattering gas density to suggest that it will generally be impossible
to observe scattered light in the vicinity of a cD galaxy.  This
example, M87, has a maximum mass inflow rate of $40 \msun {\rm
yr}^{-1}$.  Here $n_c$ is based on the cooling flow model fit to
Einstein X-ray data (White and Sarazin 1988).  Oemler (1976) gives the
core radius of 2.2~kpc and we match the number density of the cooling
flow model there.  The result is $n_c=0.75 \; {\rm cm}^{-3}$.  The
luminosity density is calculated in the same manner as above using eq.
(\lumden), for an absolute visual magnitude of -23.53 (Oemler 1976).
We find $\rho_{L_V}(0) = 4.6\times 10^{-23} \; {\rm erg \, s^{-1}
cm^{-3}}$.

Figure 7b shows the values of 1\% of the galaxy plus sky intensity,
$I$, and $P$ for M87.  Again, the scattered and polarized intensities
lie below the 1\% level.  {\it We conclude that the intensity of
visible scattered light from a cD galaxy immersed in a cooling flow is
too small to be observed.}

\subsection c. A cD Galaxy With an Isotropic Active Nucleus \par

In this section we add an active nucleus to the cD galaxy. In our
model we assume (1) that the luminosity of the central nucleus is
equal to that of the surrounding galaxy in V; (2) that the emitted
light from the nucleus is isotropic; (3) that the nucleus is a point
source; (4) that the nuclear luminosity is constant; and (5) that
emission is unpolarized.  In the next section we consider an
anisotropic flux from the central source.  For relaxation of some of
these restrictions we refer the reader to Wise and Sarazin (1992).

In this section we will compare the scattering in U, B, and V bands.
Thus, we assume (6) the parent galaxy has a spectrum of the form,
$$
F^{gal}_{\nu} \propto \left\{
\matrix{\nu^{-2.1}, &{\rm for} \;\; \nu < 10^{14.83}{\rm Hz} \cr
\nu^{-6.0}, &{\rm for} \;\; \nu > 10^{14.83}{\rm Hz}\; ,} \right.
$$
and the active nucleus of the form
$$
F^{agn}_{\nu} \propto \left\{ \matrix{\nu^{-1}, \;\; {\rm for} \;\;
\nu < 10^{14.85}{\rm Hz} \cr \nu^{0}, \;\; {\rm for} \;\; \nu >
10^{14.85}{\rm Hz} \; .} \right.
$$
These spectral distributions are rough fits to the ``standard galaxy''
taken from Malkan and Oke (1983) and to AGN's taken from Francis
(1993).  The U, B, and V intensities are determined by integrating the
above distributions over the appropriate frequency ranges.

We choose the same parameters as those for NGC~1275 except that we
include a central point source of equal luminosity, $L_V^{agn} =
1.54\times 10^{45}$ erg s$^{-1}$.  Far outside the core of the
electron scattering distribution, eq. (\Itot) implies that the
scattered intensity from the active nucleus goes as $1/r^2$. Figures 8
show the total and polarized intensities for V, B, and U.  For U and B
we take the sky intensity to be 22.2 and $23.0\; {\rm mag\;
arcsec}^{-2}$, respectively (Neizvestnyi 1982; Pilachowski {\it et
al.} 1989).  In V the total and polarized intensities exceed the 1\%
background level for radii less than 15~kpc and 4~kpc, respectively.
Moving to U these radii become 19~kpc and 13~kpc, respectively.
Within 10~kpc the measured polarization, $P$, exceeds several percent
in U. This change is largely due to the decreased galaxy background.

For all three bands, the detectable effect is due completely to the
presence of the active nucleus.  Obviously, if the nucleus were
brighter than the value we assigned it, the values of $I$ and $P$
would be proportionally larger.  Active galaxies can have luminosities
up to and exceeding $10^{47} {\rm erg\; s}^{-1}$ (Weedman 1986),
roughly 100 times the luminosity of the most luminous cD galaxy.  At a
luminosity of $10^{47} {\rm erg\; s}^{-1}$ the scattered $I$ and $P$,
for parameters similar to NGC~1275, could be detected out to roughly
200~kpc.  Although a correlation exists between the luminosities of
the active nucleus and the parent galaxy, we note that we have already
adopted a near-maximum estimate for the luminosity of the cD (Hutchings,
Crampton, and Campbell 1984) whose background serves to obscure the
scattered light.  Many more galaxies may be visible in scattered
light; this depends upon how many have nuclei that outshine the rest
of the galaxy and, at the same time, have sufficiently high Thomson
optical depths.

\subsection d. Active Nucleus with Beamed Radiation \par

There have been recent suggestions that the brightest active galactic
nuclei are those that are being beamed toward us (Barthel 1989),
though there is some evidence against this suggestion (Boroson 1992).
With this in mind we now investigate a model that has its radiation
beamed at an angle $\phi$ relative to the observer's line of sight.
Let $L_b$ be the apparent luminosity for an observer when the beam is
aligned with the line of sight.  Relative to the isotropic case, with
$L=L_b$, the intensity of scattered light will be diminished since the
emissivity is confined to the cone of radiation.  From eqs. (\Ieqns)
approximate expressions for the intensity of scattered and polarized
light are $$ I \simeq {I_0\over p^2} \Delta\theta\sin\phi (1 +
\cos^2\phi) \eqno(\neweq) $$ and $$ P \simeq {I_0\over p^2}
\Delta\theta\sin\phi (1 - \cos^2\phi) \eqno(\neweq) $$ Here $p$ is the
impact parameter, $\Delta\theta$ is the angular width of the beam, and
$I_0 = 3 L_b n_{c} \sigma_T r_c${\hfill\break}$/ 64 \pi^2$. These
formulae are accurate when the beam is not pointed near the line of
sight and $\Delta\theta$ is small.  As an example assume the gas
distribution of NGC~1275, $L_b= 10^{47} \; {\rm erg\; s^{-1}}$ and
$\Delta\theta = 15^o$.  If the beam is perpendicular to the line of
sight, $ I \simeq 9.0\times 10^{-13} / p^2{\rm (kpc) \; erg \;
s^{-1}}$ {\hfill\break}${\rm cm^{-2} arcsec^{-2}}$ or $18.4 + 5 {\rm
log_{10}}[ p{\rm (kpc) ] \; mag \; arcsec^{-2}}$ and $\Pi=1.0$.  Thus
the scattered radiation from the beam could be observed out to
$\sim$80~kpc before it drops below the 1\% background level.  Even if
the beam is not pointed toward us, its scattered light may be visible
and may be used to determine the orientation of the beam (Fabian 1989;
Wise and Sarazin 1992).

\subsection e. Spectral Lines \par

The presence of strong spectral lines from an AGN provides another
possible strategy for detecting scattered light.  If the brightness of
the emission line from the AGN is significantly larger than the
brightness of the continuum of the AGN plus galaxy background, then it
may be advantageous to use a filter narrower than U, B, or V centered
on the line frequency to enhance the signal-to-noise ratio of an
observation. Scattered light from regions far from the nucleus could be
more easily seen because background contamination would be reduced.

There are several complications.  First, the scattered line spreads in
frequency because the photons scatter off a hot gas, {\it c.f.}
\S~IIa. The expected temperatures in cooling flows range from $10^6$K
(inner few kpc) to $10^8$ (outermost regions) (Fabian \etal~1991) so
that a scattered optical line with a wavelength of $\sim5000\AA$ will
be smeared from $60\AA$ to $600\AA$.  If the nuclear continuum is flat
over similar bandwidths, the smearing reduces the brightness ratio of
the nuclear emission line to the nuclear continuum.

Observations of AGN's show that the strongest lines are roughly 3
times the AGN continuum.  Many of these lines such as H$\alpha$ have
velocity widths of a few thousand km~s$^{-1}$ or $\sim$~few~$\times
100\AA$ (Osterbrock and Mathews 1986).  Narrower lines such as [OIII]
are less useful: their maximum intensity per frequency interval is
comparable to H$\alpha$, but the smearing will reduce their
intensities relative to the continuum by a much larger factor.

To make an estimate we proceed as follows.  Some emission lines such
as H$\alpha$ arise almost solely from the photoionization of clouds in
the inner 100~pc of the nucleus. In section \S IIIc we have shown
that the most promising region for detecting scattered light is within
the inner 20~kpc of the nucleus.  At these radii the gas temperature
will be $\sim10^6$K and H$\alpha$ will be smeared $\sim100$\AA.  Since
the smearing is less than the intrinsic line width, the ratio of
scattered line to scattered nuclear continuum will be roughly 3. If
the background from the galaxy is roughly flat over the same
bandwidth, then the filtered intensity may be enhanced by up to a
factor of 3 compared to the broad-band case.  Thus, strong emission
lines emanating from an AGN could be marginally helpful for observing
scattered light in regions of a cooling flow outside of the nucleus.

One complication is that many cooling flows exhibit the same optical
emission lines as found in AGN's (Stockton and MacKenty 1987).  In
this case, the galactic background includes a component that is not
reduced by using a filter and the usefulness of the emission lines is
decreased.  Still, one could distinguish the origin of this emission
by polarization measurements and possibly by subtracting the narrower
background emission line of the cooling flow.

\subsection f. Off-Center Gas Distribution \par

We have shown that scattered and/or polarized light can currently be
detected in a cD galaxy with an active nucleus of luminosity roughly
equal to or greater than that of the rest of the galaxy.  In this
section we use the results from \S IId to examine the situation when
part of the scattering gas is not centered on the cD galaxy.  For a
galaxy of fixed luminosity, a scattering halo typical of a cD ($\alpha
= 1$) and a clump of fixed column density we showed that the ratio of
the clump scattered brightness to that of the halo increases with
distance from the galaxy center.  Two additional factors must be taken
into account to assess the visibility of scattering from such
off-center clumps: (i) the diffuse background of the galaxy and (ii)
the intrinsic sky level.

A possible example of an off-center density distribution is the radio
source PKS~0745-191 (Fabian {\it et al.}~1985). This galaxy is
immersed in a hot gas that is cooling at a rate of up to 1000 $\msun
{\rm yr}^{-1}$.  The X-ray image of this source extends to roughly
200~kpc with a peak of emission that lies 30~kpc from the center of
the galaxy (indicating a possible density maxima).  From the optical
observations of Fabian {\it et al.}~(1985) we arrive at the following
values for the galaxy: $r_c^{gal} = 9$ kpc, $r_{\rm cut} = 45$ kpc,
$m_V=16$, and $L_V=1.09\times10^{45}$ erg s$^{-1}$.  The cutoff radius
of the galaxy is uncertain but its actual value does not change our
results significantly.  From X-ray data the average density within 30
kpc of the intensity peak is $0.06 \; {\rm cm}^{-3}$ (Fabian {\it et
al.}~1985). We take the scaling radius of the off-center density peak
to be equivalent to the core radius of the cD galaxy, $r_c^{cl} = 9$
kpc. Then, assuming for the scattering gas a power law identical to
eq. (6) where $\alpha = 1$, we arrive at $n_c^{cl}=0.13 \; {\rm
cm}^{-3}$.  The luminosity is assumed to be emanating from a AGN for
the purpose of calculating the scattering.  (Consulting Figure 6, we
expect the errors in the total scattered intensity to be less than a
factor of two.)  For the purpose of determining the background
emission we assume the galaxy's luminosity density is extended,
following eq.  (\Lden) where $\beta = -3/2$ .

Using the information from the previous paragraph we now estimate the
intensity a distance $r_c^{cl}$ from the scattering gas center.  We
will assume that the gas center is coplanar with the galaxy center and
the clump gas has $\alpha = 1$.  We find the intensity of the
polarized emission, P, to be 27.5 magnitudes per square arcsecond,
about one magnitude brighter than the examples in \S IIIb at the
equivalent radius.  In our power-law model the intensity diverges as
the density singularity is approached ({\it c.f.} \S IId); the actual
maximum is dictated by the clump's core radius which, of course, is
unknown.  Overall, there is a region roughly 10~kpc in diameter near
the density peak with $I$ and $P$ above the 1\% background level.
Thus, we can only conclude that if the core radius $\lta 10 \kpc$
scattering may be visible.  If the gas center lies out of the plane of
the galaxy then the polarized intensity will be reduced significantly
relative to a gas center in the plane.  Thus, it may be possible to
detect relatively dense regions of hot gas residing near a luminous
galaxy.  If detected, it may be possible to determine the position
along the line of sight from the fractional polarization.

Finally, we point out that the presence of an active nucleus with
a luminosity greater than that assumed above lifts the level of
scattered light proportionally. Scattering may be used to probe the
galaxy's gas and cooling flow and so represents a potentially
complementary source of information to that provided by X-ray
observations.  In a cooling flow, for example, one would like to know
where the infalling gas is converted into stars plus the length scale
and growth rate of thermal instability. These angular scales are
typically smaller than the resolution of X-ray telescopes, but easily
within the range of optical instruments.

\section IV. Conclusions

In this paper we have outlined the nature of scattering in a hot gas
surrounding a luminous galaxy or an active galactic nucleus for a
variety of simple models. We began by considering a point-like
luminosity source embedded in scale-free electron power laws.  We
derived asymptotic estimates for the scattering intensity and
polarization fraction in the neighborhood of the density peak and the
luminosity source.  We complemented these analytic results with
numerical calculations of the scattering intensity, polarization
fraction, and angle of polarization for single and multiple point-like
luminosity sources.  The large-scale patterns illustrate the effects
of interference between sources. For an extended luminosity source we
showed that the interference decreased the polarization fraction in
the vicinity of the source.

We applied our results to investigate several potential types of
scattering in the vicinity of galaxies. For a typical cD (without an
active nuclei) the scattered galactic emission is too weak to be
detected with current instruments at optical wavelengths.  However,
those galaxies with an active nucleus and/or off-center gas are
possibilities.

An active nucleus increases the flux of scattered radiation and is so
small that interference effects are negligible.  We show that when a
galaxy has an active nucleus equal in luminosity to the parent galaxy,
the total and polarized scattered intensities in the inner 5--30~kpc
exceed the 1\% background level.  This is a conservative estimate
since the luminosity of many nuclei exceed their parent galaxies by
factors of 10 to 100; for these more luminous examples the scattered
and polarized light becomes visible on much larger size scales. For
typical galaxy and AGN spectra, it is significantly easier to measure
the scattering at short wavelengths (i.e. U).  If an active galaxy has
an obscuring dust torus or is beaming its radiation in a direction
away from the line of sight, it may be possible to determine the
orientation of the torus/beam by searching for the cone of scattered
light.  There is only modest advantage to using a narrow-band filter
centered on an emission line because the typical continuum associated
with the active nucleus is only a factor of 3 less in intensity than
the scatter-broadened emission lines.

For galaxies with off-center gas distributions, detection is easier
because of the reduced background from the galaxy at the gas center
and a high degree of polarization.  The polarization fraction reaches
a local maximum as the impact parameter approaches the off-center
distribution. In our example, PKS 0745-191, the scattered and
polarized intensity were greater than 1\% of the background intensity
of the galaxy and sky for a region roughly 10~kpc in diameter. When an
active nucleus is present both the intensity of the scattered
radiation and the size of the region grow. Hence, clumps of ionized
gas in the environment surrounding the galaxy should be detectable by
searching for regions of polarized emission.

We conclude that there are a number of promising observational
possibilities offered by polarized scattering that occurs near
galaxies, especially those with AGN's.

\vskip 1cm

\noindent{\bf Acknowledgements}

\noindent This research was supported in part by grants NSF-AST-8657467,
NSF-AST-8913112, NSF-AST-9119475, NAGW-2224. We thank G.  Djorgovski,
R. Goodrich, J. Krolik, C. Sarazin, I. Wasserman, and the referee for
useful comments.

\vfill
\page

{\section References \par\frenchspacing}

\apj Barthel, P. D. 1989/336/606/
\apjlett Boroson, T. A. 1992/399/L15/
\aj Burstein, D. and Heiles, C. 1982/87/1167/
\apj Canizares, C. R., Stewart, G. C. and Fabian, A. C. 1983/272/449/
\refpaper Chambers, K. C., Miley, G. K. and van Breugel, W.
	1987/Nature/329/604/
\apj Cowie, L. L., Hu, E. M., Jenkins, E. B. and York, D. G. 1983/272/29/
\mn Crawford, C. S. and Fabian, A. C. 1989/239/219/
\refbook di Serego Alighieri, S., Cimatti, A., and Fosbury, R.A.E 1992,
Arcetri Astrophysics Preprint no. 17./
\aj Djorgovski, S., Spinrad, H., Pedelty, J., Rudnick, L. and Stockton, A.
	1987/93/1307/
\mn Fabian, A. C. 1989/238/41p/
\mn Fabian, A. C., Arnaud, K. A., Nulsen, P. E. J., Watson, M. G.,
Stewart, G. C., McHardy, I., Smith, A., Cooke, B., Elvis, M. and
Mushotzky, R. F. 1985/216/923/
\apj Fabian, A. C., Hu, E. M., Cowie, L. L. and Grindlay, J. 1981/248/47/
\refpaper Fabian, A. C., Nulsen, P. E. and Canizares, C. R.
	1991/ Astr. Astrphys.
Rev./2/191/
\apj Francis, P. J. 1993/407/519/
\refpaper Gil'fanov, M. R., Syunyaev, R. A., and Churazov, E. M.
	1987/ Soviet Astro.
Letters/13/3/
\apj Goodrich, R. W. 1992/399/50/
\apjlett Heckman, T. M. 1981/250/L59/
\apj Hutchings, J. B., Crampton, D., and Campbell, B. 1984/280/41/
\apj Jones, C. and Forman, W. 1984/276/38/
\apj Kent, S. M. and Sargent, W. L. W. 1979/230/667/
\apj Kriss, G. A., Cioffi, D. F., and Canizares, C. R. 1983/272/439/
\apj Malkan, M. A. and Oke, J. B. 1983 /265/92/
\aj  McNamara, B. R. and O'Connell 1993/105/417/
\refbook Meiksin, A. 1990, {\it Hubble Symposium}, (P.A.S.P.
	Conference Series)./
\refpaper Miller, J. S. and Goodrich, R. W. 1988/Nature/331/685/
\refpaper Neizvestnyi, S. I. 1982/Bull. Spec. Astrophys. Obs./16/41/
\apj Oemler, A. J. 1976/209/693/
\aj  Oemler, A. J. and Tinsley, B. M. 1979/84/985/
\refpaper Osterbrock, D. E. and Mathews, W. G.
	1986/Ann. Rev. Astr. Astrphys./24/171/
\refpaper Pilachowski, C. A., Africano, J. L., Goodrich, B. D. and
Binkert, W. S. 1989/Publ. Astron. Soc. Pac./101/707/
\refbook Rybicki, G. B. and Lightman, A. P. 1979, {\it Radiative Processes
in Astrophysics}, (John Wiley: New York), p. 62, 90./
\refpaper Sarazin, C. L. 1986/Rev. Mod. Phys./58/1/
\apjlett Sarazin, C. L., O'Connell, R. W. and McNamara, B. R. 1992/397/L31/
\apj Sarazin, C. L. and Wise, M. W., 1993/411/in press/
\apj Serkowski, K., Mathewson, D. S. and Ford V. L. 1975/196/261/
\apj Stewart, G. C., Canizares, C. R., Fabian, A. C. and Nulsen, P. E. J.
	1984/278/536/
\apj Stewart, G. C., Fabian, A. C., Jones, C. and Forman, W. 1984/285/1/
\apj Stockton, A. and MacKenty, J. W. 1987/316/584/
\refbook Weedman, D. W. 1986, {\it Quasar Astronomy},
	(Cambridge University Press:
Cambridge)./
\refpaper Syunyaev, R. A. 1982/ Soviet Astro. Letters/8/175/
\apj White, III, R. E. and Sarazin, C. L. 1987/318/612/
\apj \bysame 1988/335/688/
\apj Wise, M. W. and Sarazin, C. L. 1990/363/344/
\apj \bysame 1992/395/387/

\vfill
\page

\hsize 6.5truein
\hoffset=0in
{\section Figure Captions\par\frenchspacing}

\ref
Figure 1. A photon leaves the source in the direction ${\hat k}$,
is scattered through an angle $\theta$ into the line-of-sight ${\hat
n}$.  The polarization bases of the the unscattered light, $\hat
\epsilon_1$ and $\hat \epsilon_2$ (see \S IIa), and the scattered
light, $\hat e_{\perp}$ and $\hat e_{\parallel}$, are illustrated.

\ref
Figure 2. The parallel and perpendicular bases for the scattered
light are displayed as seen by the observer.

\ref
Figure 3a. Polarization fraction (length of the line segments), the
angle of polarization (orientation of the line segments), and the
intensity of the scattered light (contours) for a luminosity source
coplanar with the gas center.  Here the luminosity source is coplanar
with and $1r_c$ to the right of the gas-density origin. The contour
interval is one-half magnitude ($10^{0.2}$).

\ref
Figure 3b. Contours of polarization fraction are shown for the
same geometry as in Figure 3a.  The contour interval is .04.

\ref
Figure 3c.  Same as 3a but the luminosity source is $4r_c$ out of
the plane of the gas-density origin.

\ref
Figure 4a.  Same as 3a but with two luminosity sources coplanar with
the gas-density origin.

\ref
Figure 4b.  Same as 3a but with two luminosity sources, one coplanar
with the gas-density origin and one $4r_c$ out of the plane.

\ref
Figure 5.  Polarization as a function of radius for an extended source
given by eq (\exttwo).

\ref
Figure 6. The ratios of the extended to point source scattered
intensities for $I$ and $P$.

\ref
Figure 7a. The scattered brightness profile of NGC~1275 ($I$ and $P$)
is compared to the threshold for detection, which is 1\% of the sum of
the extended diffuse galactic emission and the sky.  $I$, $P$, and the
1\% background level are represented by the dashed, dotted, and solid
lines, respectively.  At no point is the scattering detectable in
optical wavelengths with current instruments (see \S IIIb).

\ref
Figure 7b. Same as 7a for M87.

\ref
Figure 8.  Same as 7a except that a central point source of luminosity
$L_V=1.54\times 10^{45} {\rm erg\,s^{-1}}$ (equal to that of the
background galaxy) is included in the calculation for three wavelength
bands, U, B, and V. $I$, $P$, and the 1\% background level are
represented by the dashed, dotted, and solid lines, respectively.
Note that $I$ and $P$ exceed the detection threshold for this case.

\vfill
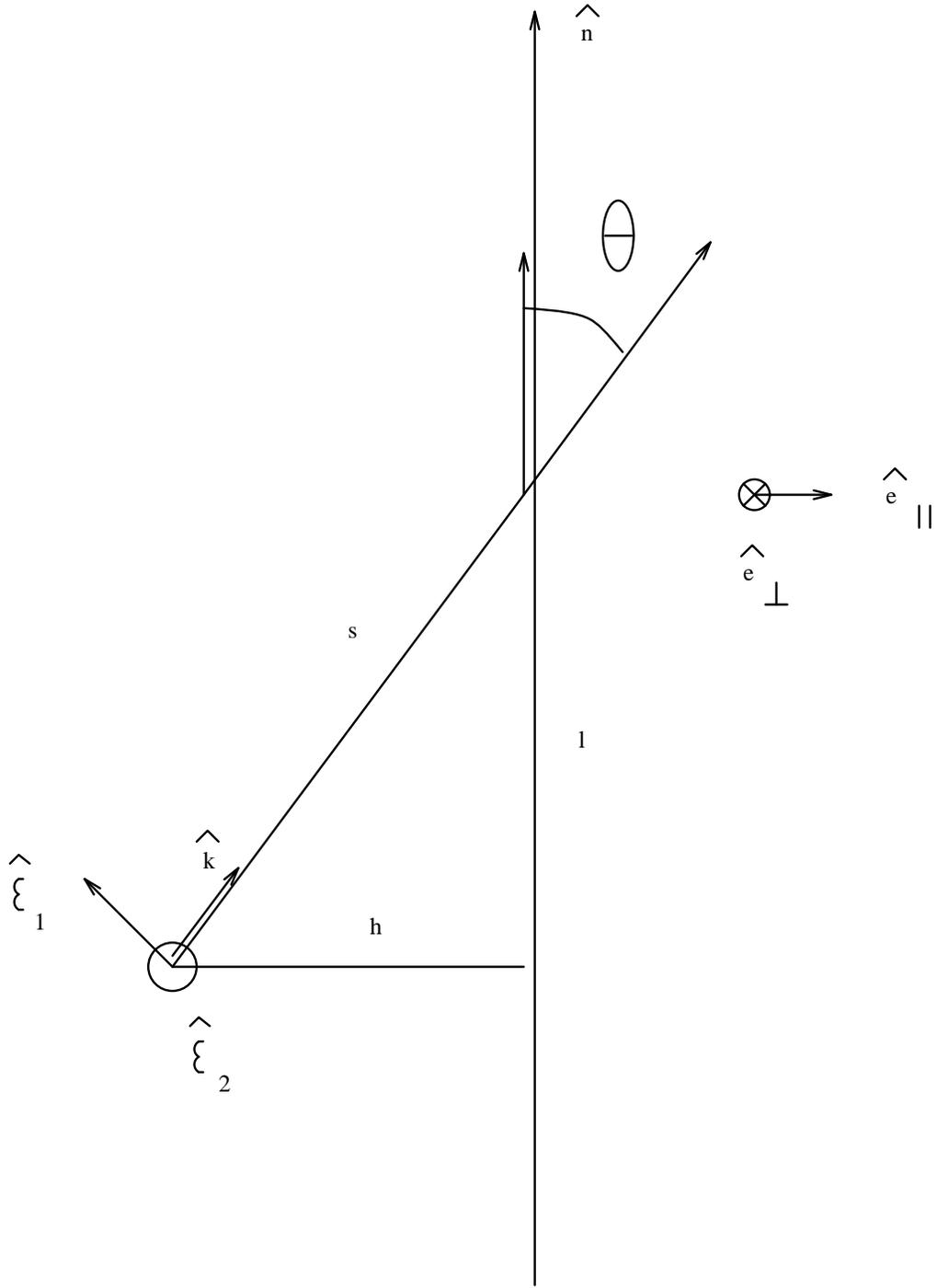
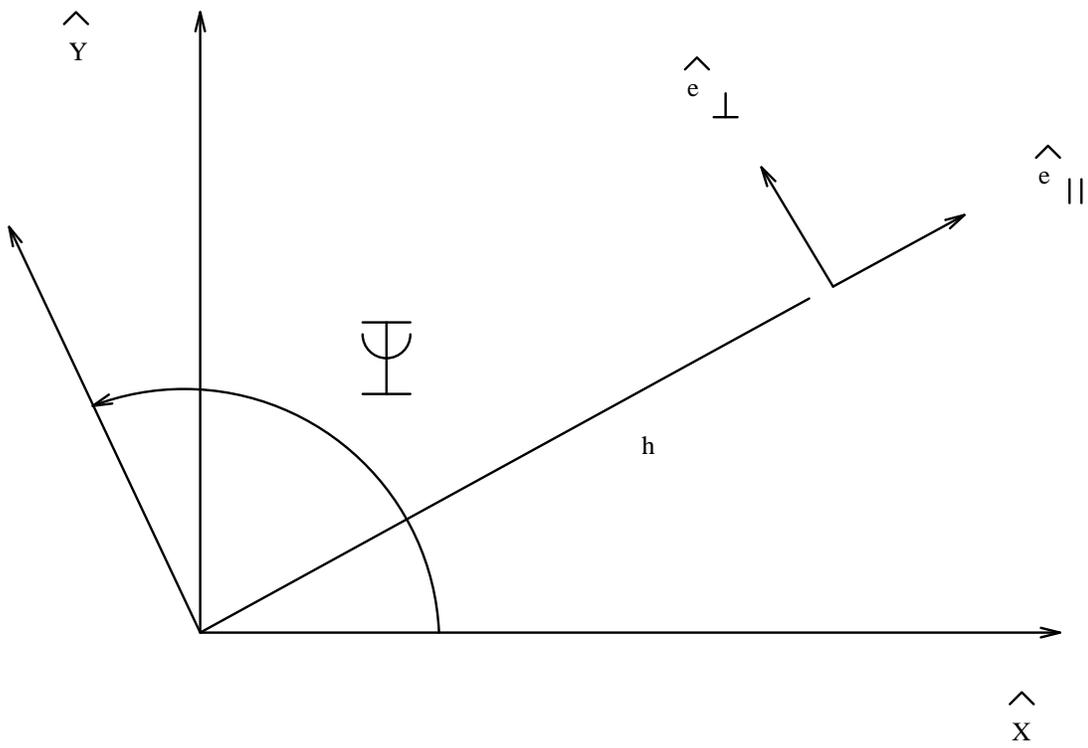
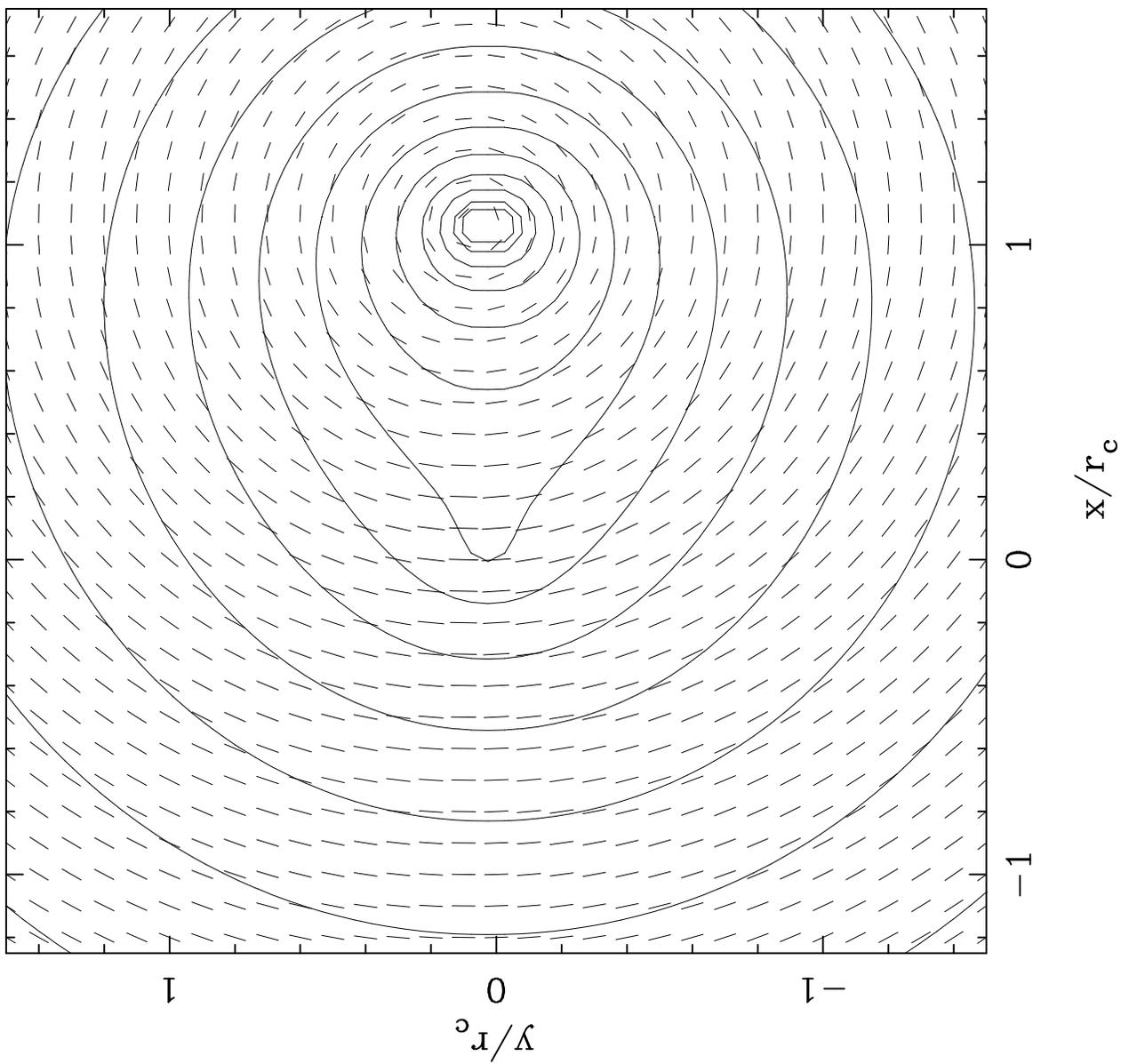
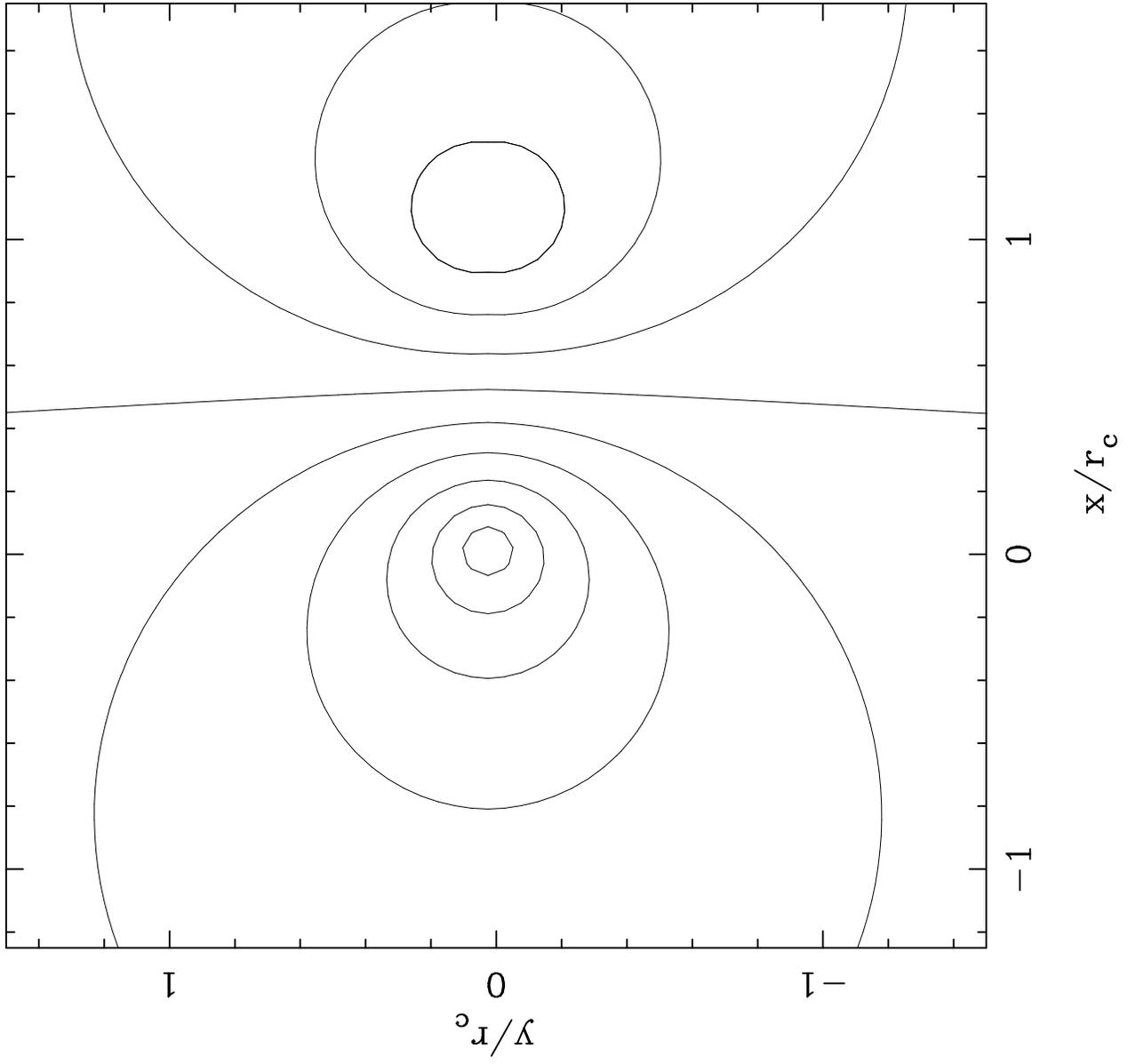
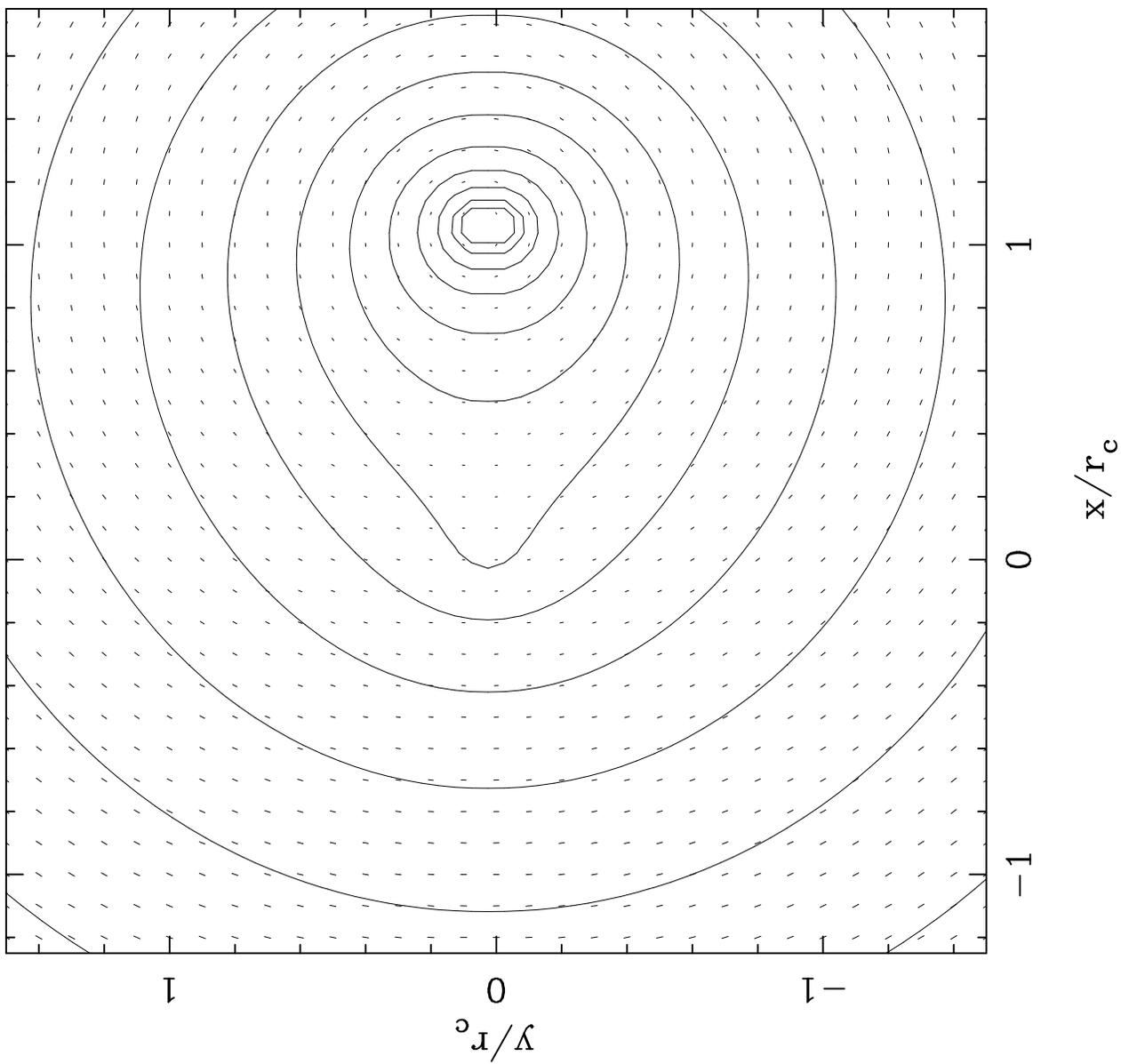
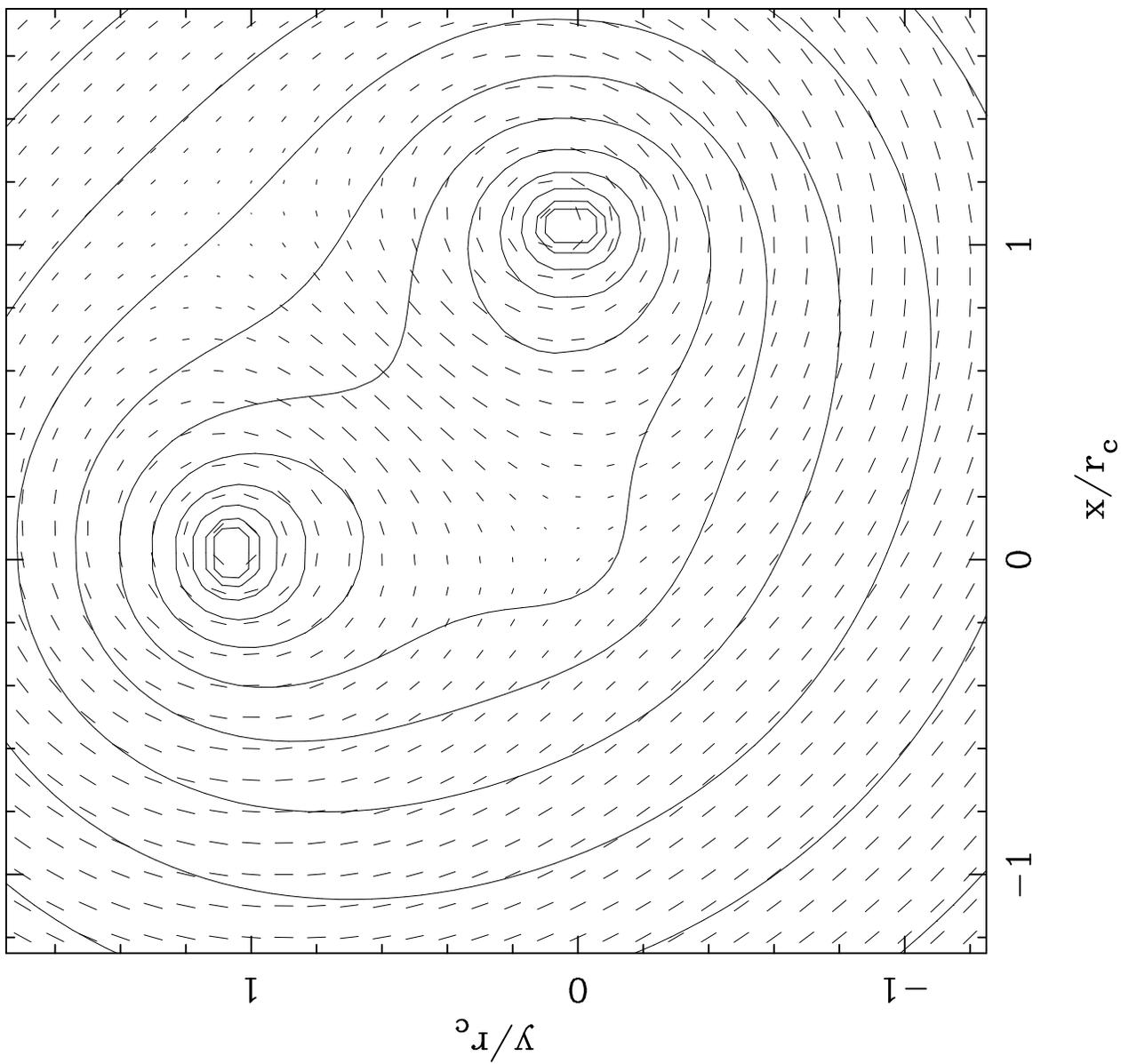
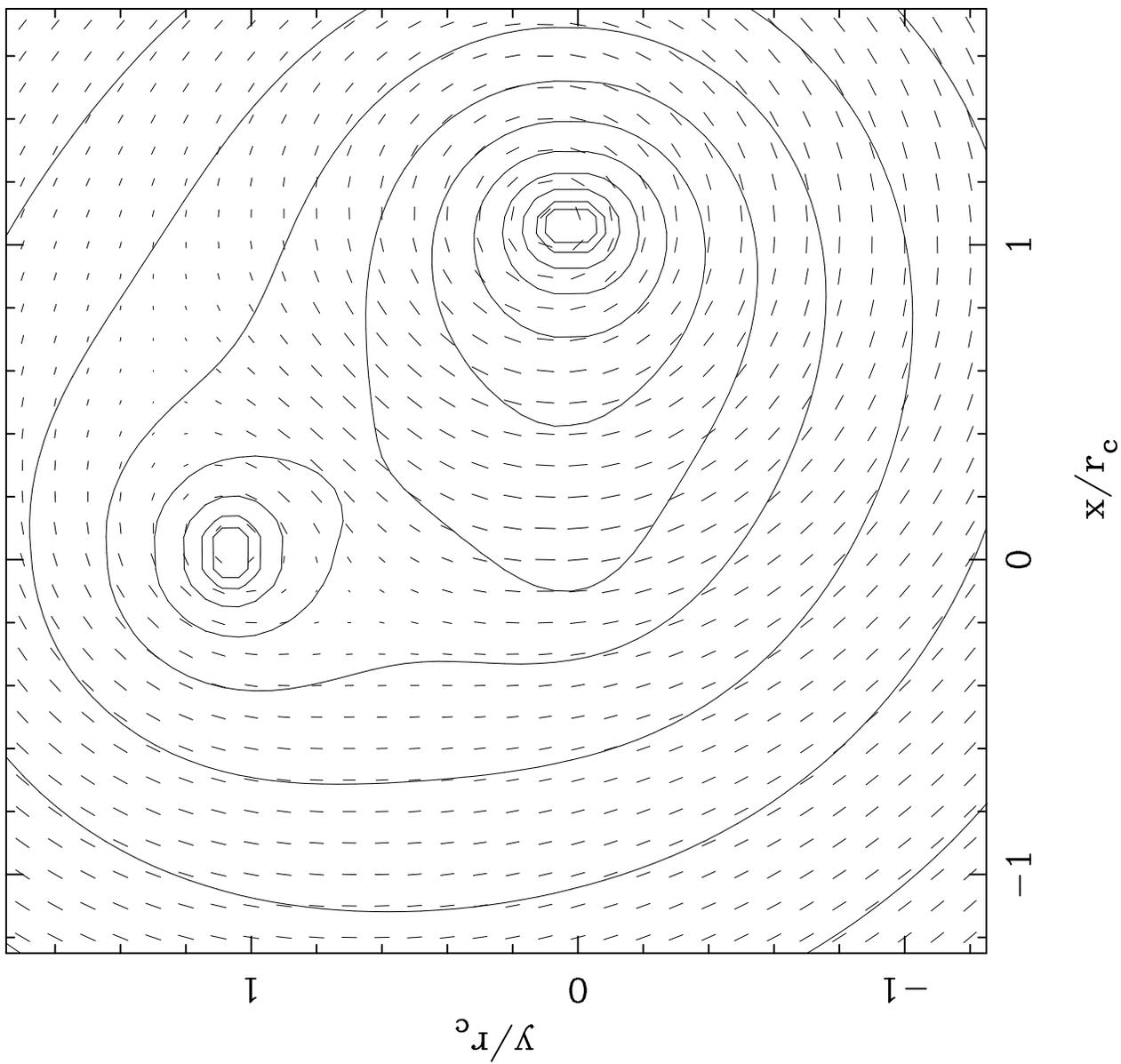
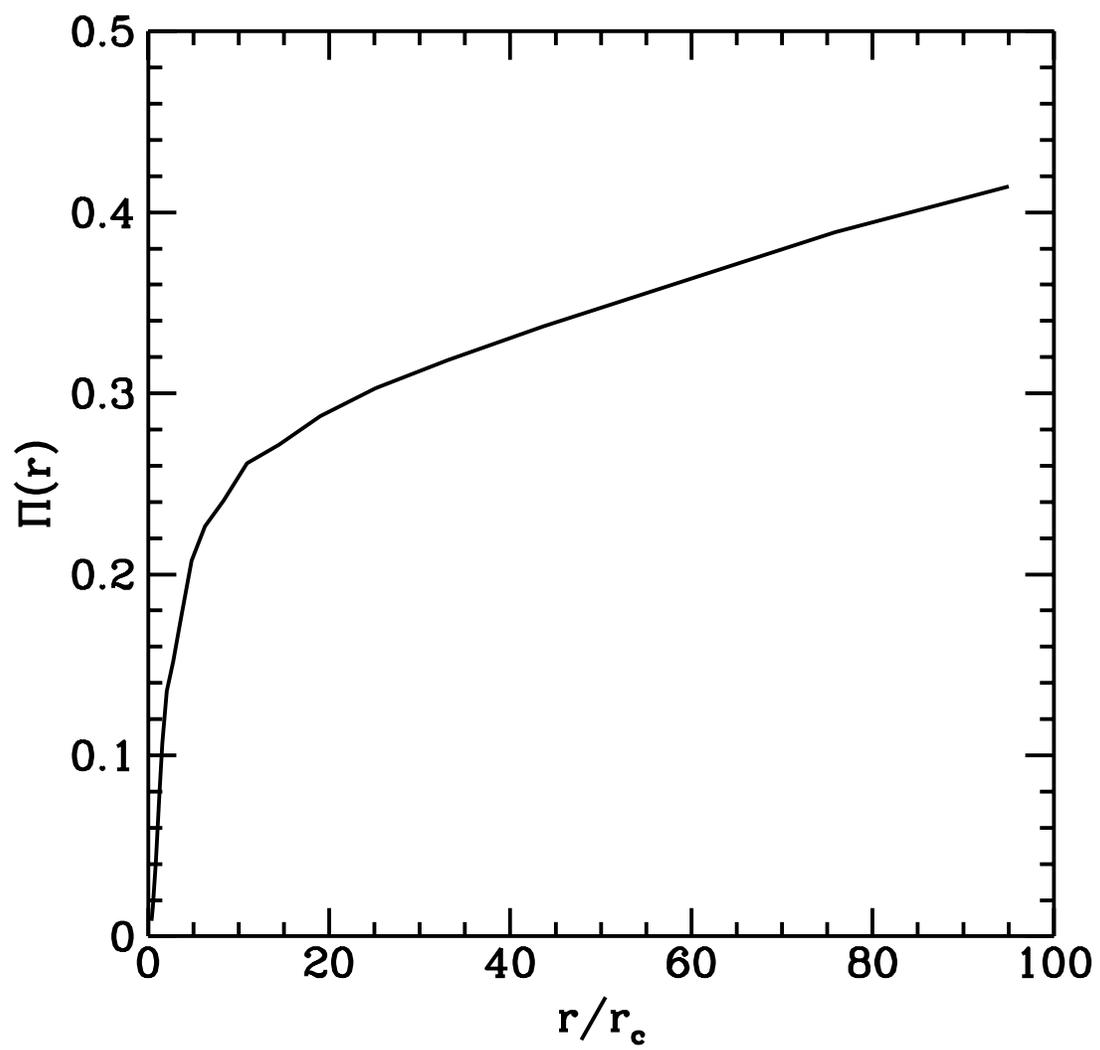
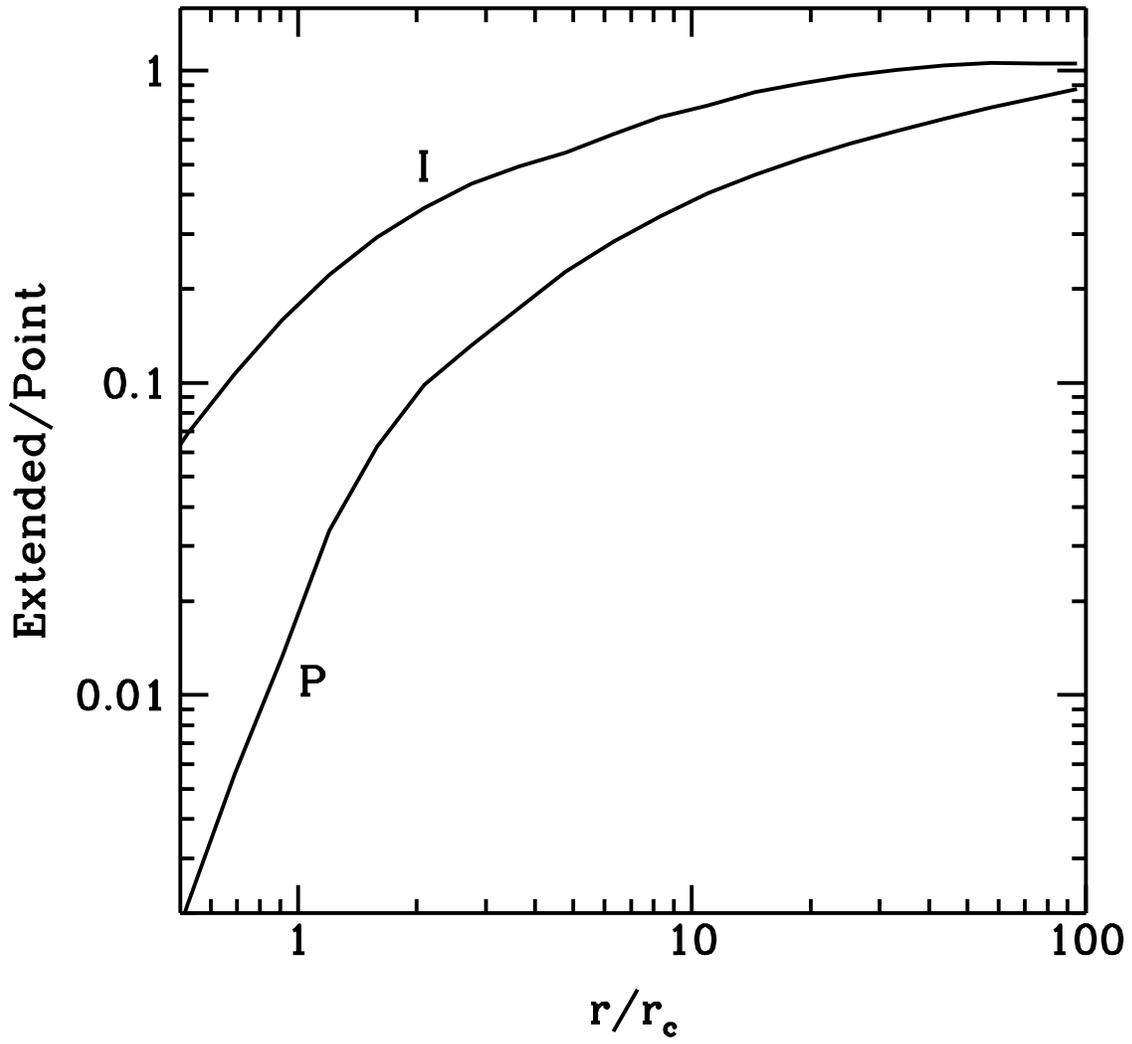
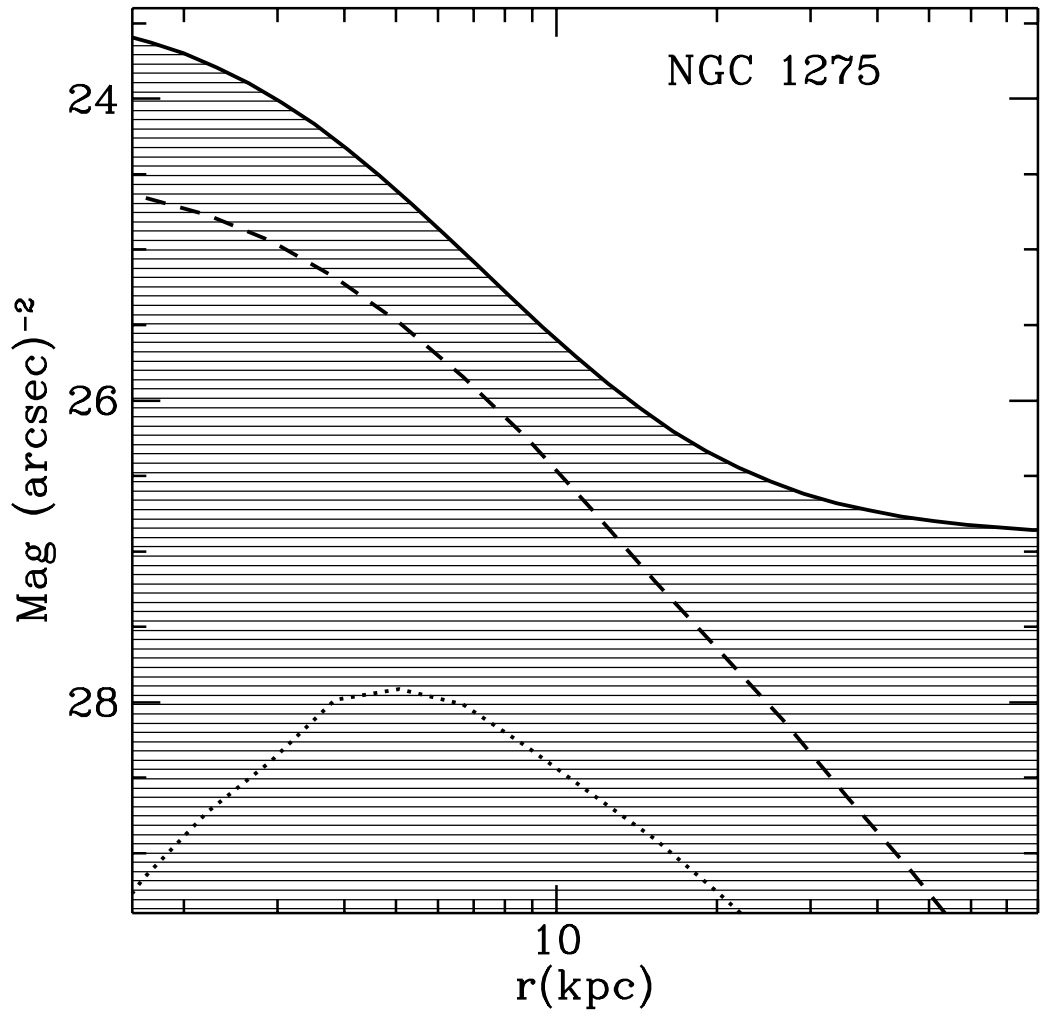
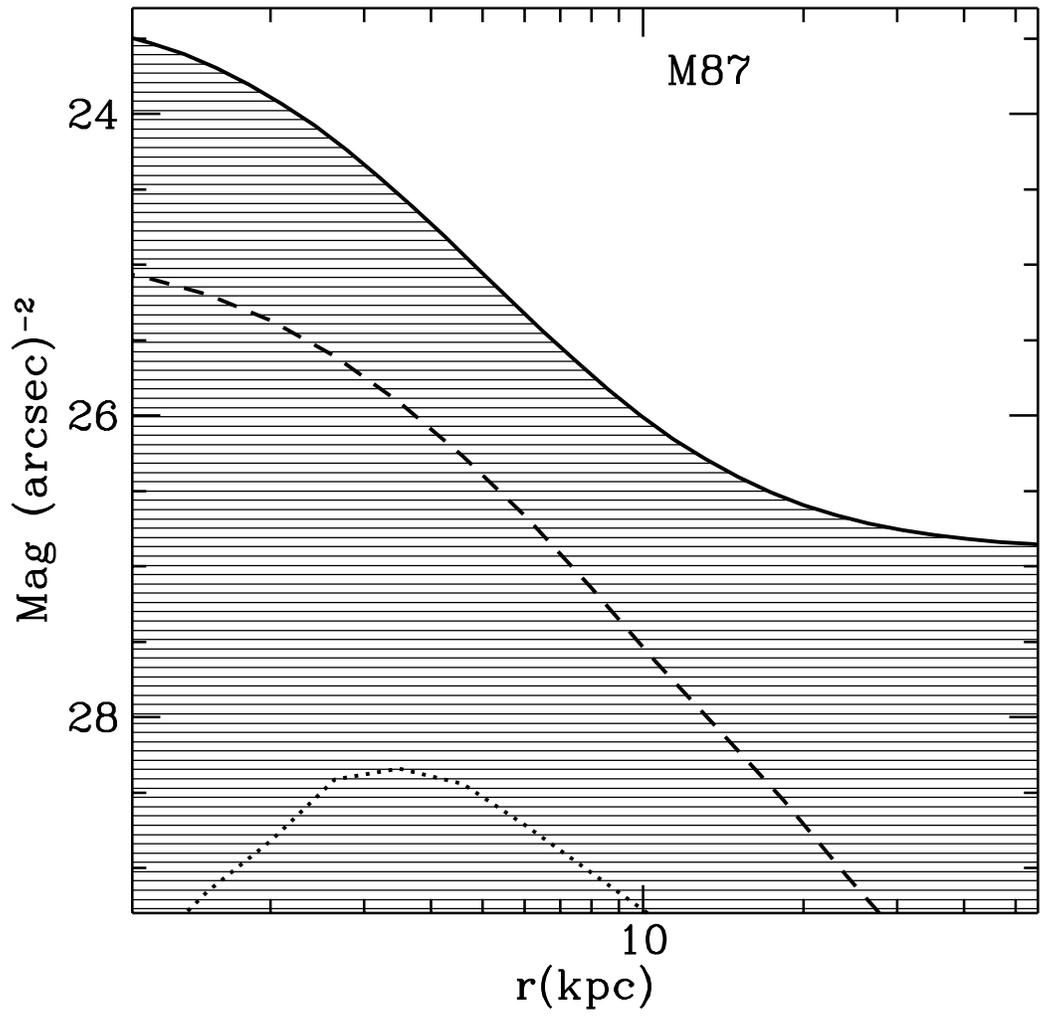
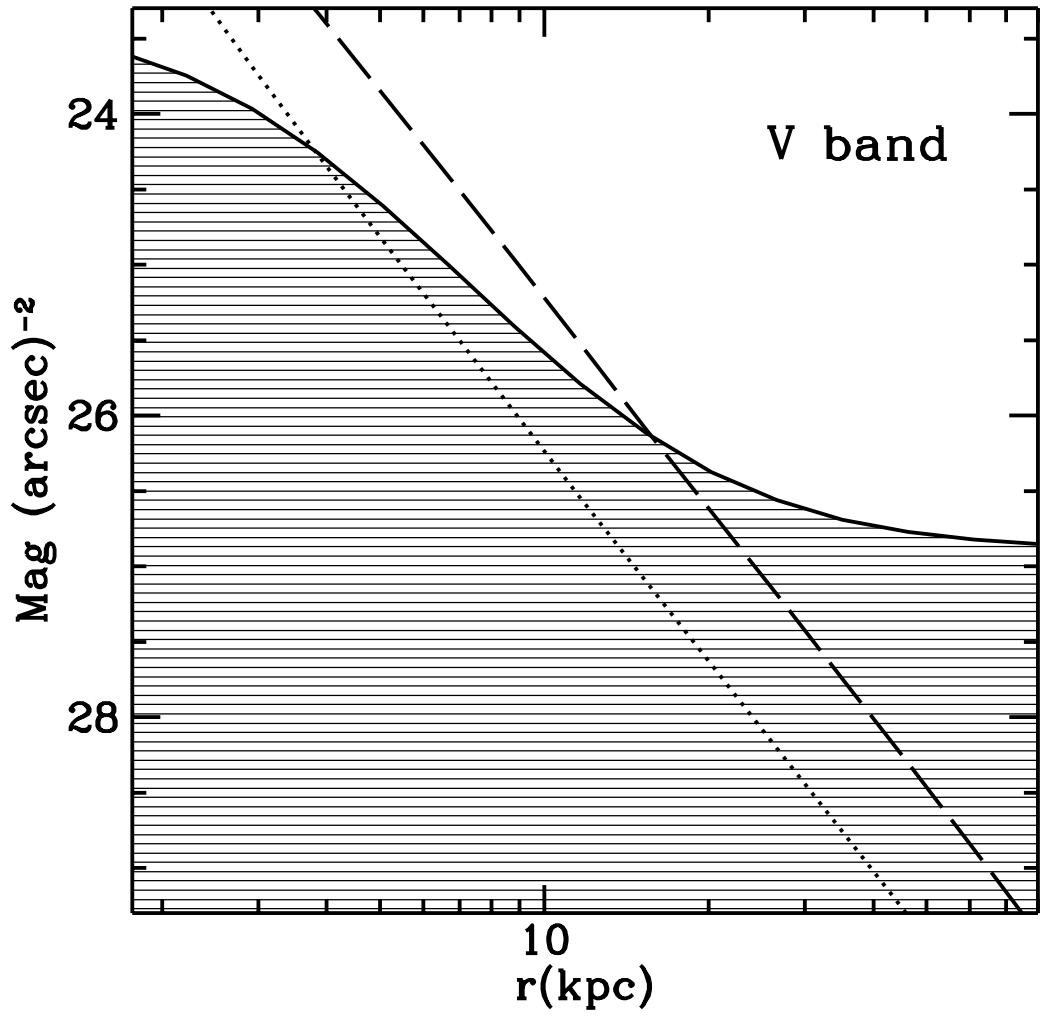
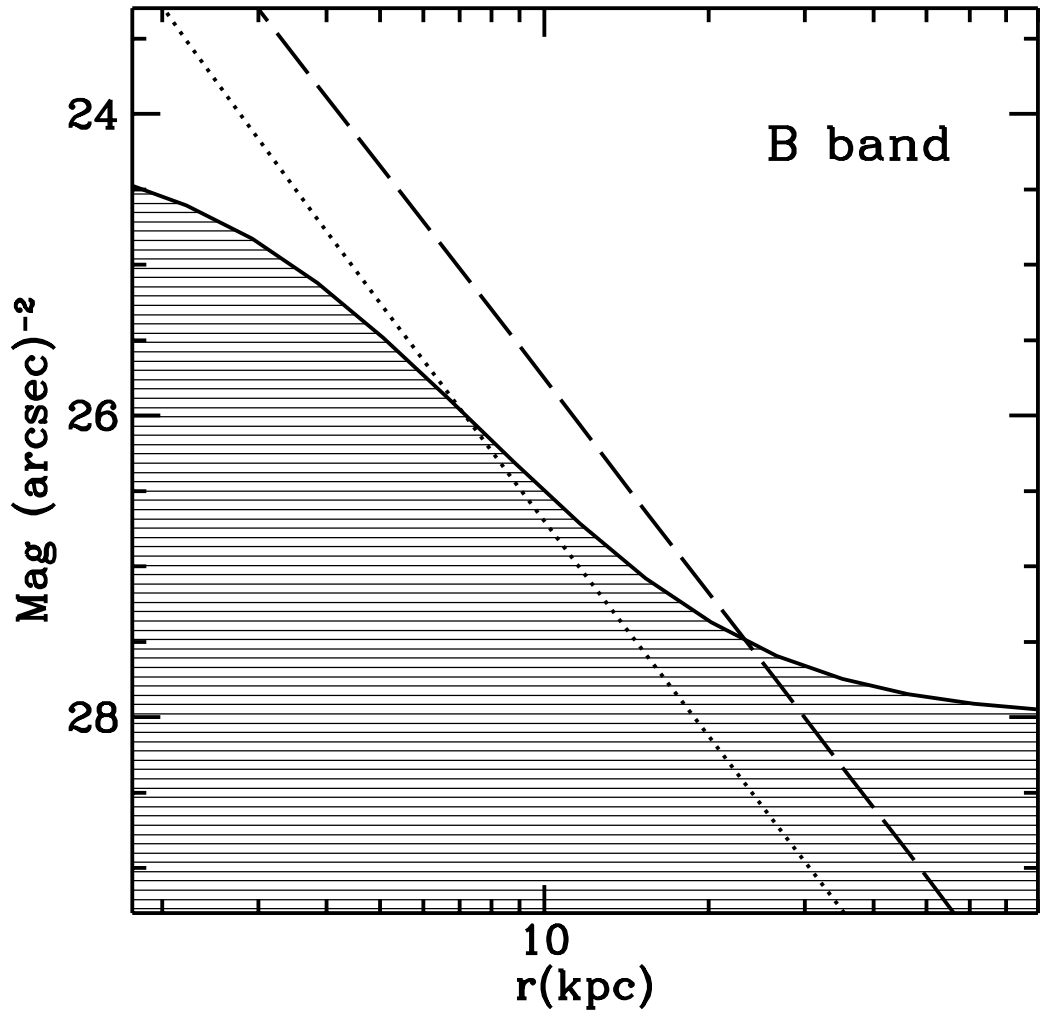
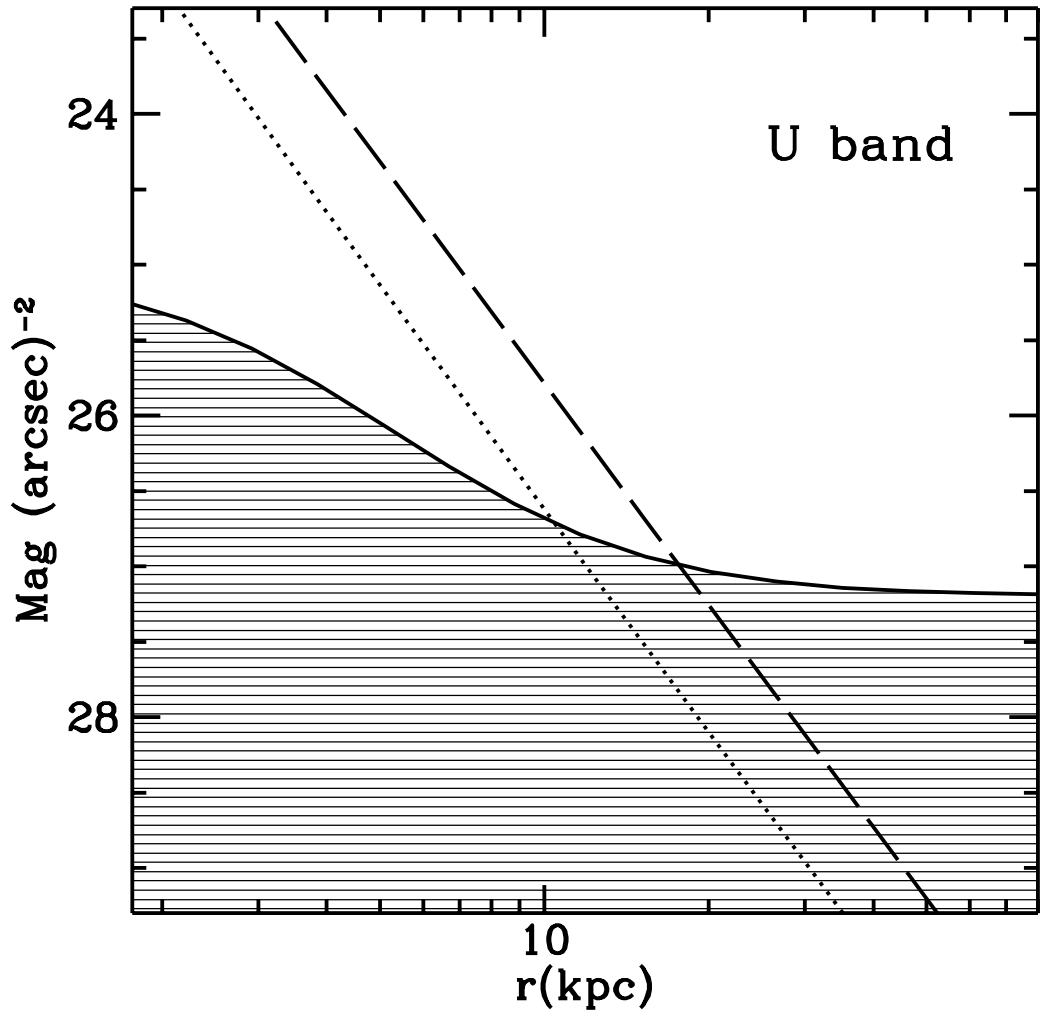
\end